\documentclass[prl,twocolumn,notitlepage,superscriptaddress,longbibliography]{revtex4-1}
\usepackage{amsmath}
\usepackage{amssymb}

\usepackage[unicode=true,colorlinks=true,citecolor=blue,urlcolor=blue]{hyperref}

\usepackage{bm}
\usepackage{color}
\usepackage{epsfig}

\usepackage[normalem]{ulem}

\renewcommand {\phi}{{\varphi}}
\newcommand {\rmi}{{\rm i}}

\newcommand {\e}{{\rm e}}

\begin{document}
\title{
%Many-photon entangled states from a $\Lambda$-type atom in a waveguide 
Many-photon scattering and entangling in a waveguide with a $\Lambda$-type atom
%Entangling $\Lambda$-type atom with many photons in a waveguide
%(tentative titles:)\\
%Emission of many-photon entangled states by a $\Lambda$-atom in a waveguide \\
%Waveguide QED with $\Lambda$-atom for generation of many-photon entangled states\\
}

\author{Denis Ilin}
\affiliation{Department of Physics and Technology, ITMO University, St. Petersburg, 197101, Russia}

\author{Alexander V. Poshakinskiy}
%\affiliation{Ioffe Institute, St. Petersburg 194021, Russia}
\affiliation{ICFO-Institut de Ciencies Fotoniques, The Barcelona Institute of Science and Technology, 08860 Castelldefels (Barcelona), Spain}

\begin{abstract}
We develop the analytical theory that describes simultaneous transmission of several photons through a waveguide coupled to a $\Lambda$-type atom. We show that after transmission of a short few-photon pulse, the final state of the atom and all the photons is a genuine multipartite entangled state belonging to the $W$ class. The parameters of the input pulse are optimized to maximize the efficiency of three- and four-partite $W$-state production. % that reaches $?$.     
\end{abstract}

\date{\today}

\maketitle 
\twocolumngrid

{\it Introduction.}
Generation of entangled states is of paramount importance for modern quantum technologies~\cite{Ghne2009,Horodecki2009}. Two-photon entangled Bell states are the basis of quantum communication.  Multipartite entanglement is harder to achieve as is requires all the particles to interact. However, it promises strong benefits, e.g.,  Greenberger--Horne--Zeilinger (GHZ) states can be used for superdense coding and quantum teleportation between several parties~\cite{Karlsson1998}. $W$-states~\cite{Dur2000} are also suitable for these tasks~\cite{Agrawal2006} but, in contrast to GHZ states, their entanglement is robust against the loss of one of the particles.  Multipartite entangled cluster states~\cite{Briegel2001} can implement measurement-based quantum computing~\cite{Raussendorf2001}.

%Of special interest are entangled states of photons as they a 
Polarization-entangled photons can be obtained with  linear optics elements only using Knill--Laflamme--Milburn (KLM) protocol~\cite{Knill2001}. However, such schemes require post-selection and usually have quite small success probability, which makes them hardly suitable for generation of multipartite entanglement. 
A promising way-around that avoids post-selection is to use quantum objects that have strong nonlinear optical properties even at a few-photon scale. As such, waveguide quantum electrodynamic (WQED) setups with natural or artificial two-level atoms strongle coupled to waveguides~\cite{Roy2017,Chang2018,Sheremet2023} can perform, e.g., nonlinear-sign (NS) gate~\cite{Ralph2015} and obtain entangled photons in single-rail encoding. 
%for two frequency-encoded qubit which can be used to obtain entangled photons.
A modulated system can generate entanglement for frequency-bin photonic qubits~\cite{Ilin2023}. 

\begin{figure}[b]
\centering
\includegraphics[width=0.7\columnwidth]{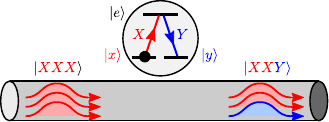}
\caption{Schematics of the photon transmission through a waveguide coupled to a $\Lambda$-type atom.  If the atom in $|x\rangle$ state is excited by several $X$-polarized photons, one of them can be converted to $Y$ polarization.  
}\label{fig:1}
\end{figure}

Atoms with more complicated level schemes offer more opportunities for entanglement generation. Consider a three-level $\Lambda$-type atom with two ground states, $|x\rangle $ and $|y \rangle$ and a single excited state $|e \rangle$, where the transitions between the $|x(y)\rangle$ state and the $|e\rangle$ state are induced by $X(Y)$-polarized photons [Fig.~\ref{fig:1}]. Such setup enables a single-photon Raman interaction (SPRINT) -- a process when an $X$-polarized photon after scattering by the atom in the $|x\rangle$ state becomes $Y$-polarized and the atom switches to the $|y \rangle$ state~\cite{Pinotsi2008,Rosenblum2015}. This allows to realize the SWAP operation between the states of the atom and a single photon~\cite{Koshino2010,Rosenblum2017,Bechler2018}. A multi-step protocol to
entangle a train of single photons was also proposed~\cite{Aqua2019}.
Systems with more complicated four-level schemes, e.g., a quantum dot in an external magnetic field, were demonstrated to generate linear polarization-entangled photonic clusters~\cite{Lindner2009,Schwartz2016,Istrati2020}.  Long-lived quantum correlations of photons also arise in such system~\cite{Smirnov2017}. 
To generate 2D cluster states, waveguide and cavity QED setups were proposed~\cite{Pichler2017,Thomas2022}.

Importantly, all previous proposals for few-photon entangling were multi-step protocols
were certain operations are performed between the subsequent emission of entangled photons~\cite{Lindner2009,Pichler2017,Aqua2019,Thomas2022}. 
Here, we show how many-photon entangled states can be generated in a single shot by a $\Lambda$-type atom in a waveguide, see Fig.~\ref{fig:1}. 
We consider the atom in the $|x\rangle$ state that is excited by a few-photon short $X$-polarized pulse. Note that upon transmission at most one of the photons can switch from $X$ to $Y$ polarization~\cite{Rosenblum2015}. Indeed, after emission of the $Y$-polarized photon, atom gets to the $|y\rangle$ state and does not interact with the subsequent $X$-polarized photons. As this conversion can happen to any one photon, the final state appears to be a polarization-entangled state of all the photons and the atom.

%as sketched in Fig.~\ref{fig:1}. The atom has two ground states, denoted $|x\rangle $ and $|y \rangle$ and a single excited state $|e \rangle$. The transitions between the $|x(y)\rangle$ state and $|e\rangle$ state can be induced by an $X(Y)$-polarized light in the waveguide.

%%% More oppotuninies -- many level. Lambda SPRINT

%
%A more complicated quantum objects with four or more levels can be used to generate
%%like charged quantum dots subjected to external magnetic field can generate on-demand
% polarization-entangled linear cluster states~\cite{Lindner2009}, or 2D cluster states in single-rail encoding~\cite{Pichler2017}. 
%
%
%%%%%%%%%%%%%%%%%%%%%%%%%%

While the essence of the effect is quite intuitive, the calculation of the scattering matrix for more than one photon is a complicated problem. For two-level atoms, the few-photon scattering amplitude was first calculated in Ref.~\cite{Yudson1984} using Bethe ansatz. Then, the result for two photons was reproduced by several other simpler methods~\cite{Fan2007,Shi2009,Baranger2013}, and the quite general theories of many-photon scattering based on input-output formalism~\cite{Xu2015,Trivedi2018} and master-equation approach~\cite{Shi2015} were developed. 
% For a $\Lambda$ atom the problem is complicated by the fact the atom can reside in the two ground states, making some of the approaches inapplicable and others cumbersome.
Three-level $\Lambda$-type atoms were often exploited in schemes where only one of the transitions is coupled to waveguide photons and the other is driven by external field leading to quantum correlations in the emission~\cite{Tudela2015,Caneva2015,Fang2016,Iversen2021}. 
However, to entangle photons, there must be at least two orthogonal photonic states, thus two active transitions are required. 
For  $\Lambda$-type atoms where both transitions are coupled to waveguide photons, only the scattering of single photons~\cite{Witthaut2010,Li2012,Bradford2012,Martens2013,Das2018,Zhong2023,Zhang2023,Chan2023}, or trains of single photons with large delay~\cite{Aqua2019}, was considered up to now. Using the the diagrammatic approach, we develop here the analytical theory that gives simple  explicit expressions  describing simultaneous transmission of $2$ and $3$ photons and allows us to maximize the entanglement efficiency.
%We develop here an analytical theory of many-photon scattering based on the diagrammatic approach that yields simple explicit expressions that describe simultaneous transmission of $2$ and $3$ photons which.
The generalization to larger photon numbers is straightforward. 
%Our rigorous theory confirms the initial idea of many-photon entanglement generation and allows us to maximize its efficiency.

%.   !!!!!!!!!!!!!!!!!!!!!!!!!!!!!
%% Also cite: cluster generation protocol~\cite{Chan2022}, Smirnov papers -- 4 level charged QD~\cite{Leppenen2021} 

%%%%%%%%%%%%%%%%%%%%%%%%%%
\begin{figure}[t]
\centering
\includegraphics[width=0.99\columnwidth]{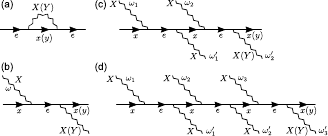}
\caption{Diagrammatic representation of (a) self-energy of the excited state, (b,c,d) non-trivial contributions to the transmission amplitude of one, two, and three $X$-polarized photons through the atom which is initially in the $|x\rangle$ state. Thin  and thick solid lines denote the Green functions of the atom in the ground state $G_{x(y)}(\varepsilon)=1/(\varepsilon-\varepsilon_g+\rmi 0)$ and in the excited state $G_{e}(\varepsilon)=1/(\varepsilon-\varepsilon_e+\rmi \Gamma_0)$, respectively. Wavy lines denote photons in the waveguide, the vertices correspond to the coupling constant $g$. 
}\label{fig:dia} 
\end{figure}
%%%%%%%%%%%%%%%%%%%%%%%%%%

{\it Model.} 
A $\Lambda$-type three-level atom coupled to a waveguide mode is described by the Hamiltonian
\begin{align}
H &= \sum_k \omega_k (a_{k,x}^\dag a_{k,x} + a_{k,y}^\dag a_{k,y})  + \varepsilon_g (b_x^\dag b_x + b_y^\dag b_y)  \nonumber\\ &+ \varepsilon_e b_e^\dag b_e
+ g \sum_k (a_{k,x} b_e^\dag b_x + a_{k,y} b_e^\dag b_y + {\rm H.c.}) 
\end{align}
where $a_{k,x(y)}$  are the bosonic operators for the photons, $b_{x(y)}$ and $b_e$ are the fermionic operators for the electron in the atom,  $\omega_k = c |k|$ is the phtoton dispersion that is assumed to be the same for $X$- and $Y$-polarized waveguide modes, $\varepsilon_g$ is the energy of the ground atomic states $|x\rangle$ and $|y\rangle$, $\varepsilon_e$ is the energy of the excited atomic state $|e\rangle$, and $g$ is the matrix element of dipole interaction. 
For the sake of simplicity in what follows, we focus on the chiral case, i.e., suppose that the atom interacts with the photons moving in one direction only, $k>0$. Generalization to the case of symmetric coupling is straightforward and discussed in the end of the paper.  

To describe photon scattering we use the diagrammatic approach outlined in Ref.~\cite{Sheremet2023}. First, the atomic states are dressed by interaction with the photons [Fig.~\ref{fig:dia}(a)] leading to the imaginary correction $-\rmi \Gamma_0$ to the energy of the excited state. Here, $\Gamma_0=g^2/c$ is the radiative decay rate of the excited state. 
Before considering several-photon scattering we briefly review single-photon transmission~\cite{Pinotsi2008}. We suppose that the  atom is initially in the $|x\rangle$ state. Then the $Y$-polarized photon does not interact with it while the transmission of $X$-polarized photon  is described by the diagram in Fig.~\ref{fig:dia}(b). The final state of the system reads $t(\omega)|Xx\rangle +s(\omega) |Yy\rangle$, which is a polarization-entangled state of the atom and the photon. Here we introduced the coefficients of photon transmission with and without polarization conversion,
%\begin{align} \label{Eq:s}
%s(\omega) &= - \frac{\rmi\Gamma_0}{\omega-\omega_0+\rmi\Gamma_0}\,,\\\label{Eq:t}
%t(\omega) &= 1+s(\omega) =   \frac{\omega-\omega_0}{\omega-\omega_0+\rmi\Gamma_0} \,,
%\end{align}
\begin{align} \label{Eq:st}
s(\omega) = - \frac{\rmi\Gamma_0}{\omega-\omega_0+\rmi\Gamma_0}\,,\quad
t(\omega) = 1+s(\omega)  \,,
\end{align}
where $\omega$ is the frequency of the photon and $\omega_0 = \varepsilon_e-\varepsilon_g$. 

%%%%%%%%%%%%%%%%%%%%%%%%%%
\begin{figure}[t]
\centering
\includegraphics[width=0.99\columnwidth]{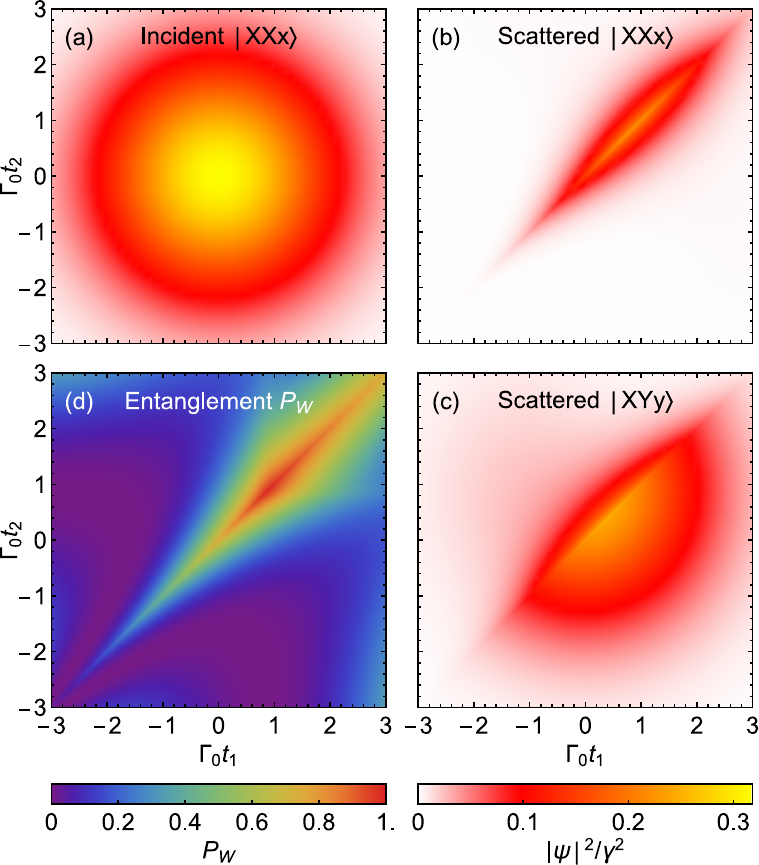}
\caption{Real-time wave functions of (a) the incident two-photon Gaussian pulse Eq.~\eqref{Eq:Gauss} with $\omega=\omega_0$, $\gamma=0.5\Gamma_0$ and (b,c) of the scattered pulse calculated after Eq.~\eqref{Eq:psi2}. (d) The conditional probability of converting the entangled state of the atom and two photons into the canonical $W$ state by SLOCC given the photons are detected at times $t_1$ and $t_2$, calculated after Eq.~\eqref{Eq:Pw}. 
}\label{fig:psi2} 
\end{figure}
%%%%%%%%%%%%%%%%%%%%%%%%%%
%
 
{\it Two-photon scattering.} 
Now we consider atom in the $|x\rangle$ state that is excited simultaneously by two $X$-polarized photons with frequencies $\omega_1$ and $\omega_2$. The nontrivial contribution to the amplitude of the process, when both photons interact with the atom, is shown in Fig.~\ref{fig:dia}(c). Additionally, the amplitudes of the processes when only one or none of the photons interacts have to be added. The scattering matrix elements corresponding to the final states $|XYy\rangle$ and $|XXx\rangle$ read
\begin{align}
&S^{XYy\leftarrow XXx}_{\omega_1',\omega_2'\leftarrow\omega_1,\omega_2} = 
(2\pi)^2 s(\omega_2') \delta(\omega_1-\omega_1')\delta(\omega_2-\omega_2')  \label{Eq:Sy} \\\nonumber
&+\frac{2\pi \rmi s(\omega_1)s(\omega_2')}{\omega_1-\omega_1'+\rmi 0}    \delta(\omega_1+\omega_2-\omega_1'-\omega_2') + (1\leftrightarrow 2) \,, \label{Eq:Sx}\\
& \nonumber \\
&S^{XXx \leftarrow XXx}_{\omega_1',\omega_2'\leftarrow\omega_1,\omega_2} = S^{XYy\leftarrow XXx}_{\omega_1',\omega_2'\leftarrow\omega_1,\omega_2}+ S^{XYy\leftarrow XXx}_{\omega_2',\omega_1'\leftarrow\omega_1,\omega_2} \\\nonumber
&+ (2\pi)^2 [\delta(\omega_1-\omega_1')\delta(\omega_2-\omega_2')+ \delta(\omega_1-\omega_2')\delta(\omega_2-\omega_1')]  \,.
\end{align}
%In the scattering amplitudes, one can separate the coherent part, which conserves frequencies of each photon, and the incoherent part, that conserves the total energy only~\footnote{see Supplementary for explicit expressions}.  
%
%The amplitude $S_{XXx,XXx}$ is symmetrized with respect to the frequencies of final photons $\omega_1',\omega_2'$, while $S_{XYy,XXx}$ is not because the final photons have different polarization. 
%
The probabilities that the atom will end in $|x\rangle$ or $|y\rangle$ state are determined by the integrals  over the final frequencies of $|S^{XXx\leftarrow XXx}|^2$ and $|S^{XYy\leftarrow XXx}|^2$, respectively. However, the latter integral of $|S^{XYy\leftarrow XXx}|^2$ has non-integrable singularities at $1/(\omega_{1(2)} - \omega_1')^2$. Interestingly,  for  $S^{XXx\leftarrow XXx}$
%, which is additionally symmetrized with respect to the frequencies of final photons $\omega_1',\omega_2'$, 
that singularities vanish and the integration result is finite~\footnote{See Supplementary for for the simplified expression for $|S^{XXx\leftarrow XXx}|^2$ that matches the well-known result for a two-level atom~\cite{Fan2007}}. 
This mathematical observation has a clear physical meaning: When excited by monochromatic $X$-polarized light, the atom with the dominant probability switches to the $|y\rangle$ state. Indeed, as the light drives the $|x\rangle \to |e\rangle$ transition only and the relaxation from $|e\rangle$ goes to both $|x\rangle$ and $|y\rangle$ states, the atom will eventually relax to the $|y\rangle$ state and stay there forever. 

%Figure~\ref{fig:3} shows the color plots of the incoherent scattering intensity, that is the squared amplitude of the part of the scattering matrix that   

To avoid the above-mentioned singularity, excitation by pulses with finite duration rather than monochromatic continuous waves should be considered. 
The wave function of the incident state for the pulse consisting of two identical $X$-polarized photons reads
\begin{align}
\psi^{\rm (in)}_{t_1,t_2} = \phi^{(0)}_{t_1}\phi^{(0)}_{t_2} |XXx\rangle \,,
\end{align}
where $\phi^{(0)}_{t}$ is the pulse envelope, $\int |\phi^{(0)}_{t}|^2 dt = 1$. 
The transmitted pulse is then described by the wave function
\begin{align}\label{Eq:psiout}
\psi^{\rm (out)}_{t_1,t_2} = \psi^{XXx}_{t_1,t_2} |XXx\rangle %\\\nonumber
+
\psi^{XYy}_{t_1,t_2} |XYy\rangle +
\psi^{XYy}_{t_2,t_1} |YXy\rangle ,
\end{align}
where
\begin{align}
\psi^{XXx}_{t_1,t_2} &= \phi^{(\tau)}_{t_1}\phi^{(\tau)}_{t_2} -[\phi^{(s)}_{t_<}]^2 \e^{-(\rmi\omega_0+\Gamma_0)|t_2-t_1|} , \label{Eq:psi2} \\\nonumber
\psi^{XYy}_{t_1,t_2}& = 
%\theta(t_2-t_1)\left[\phi^{(\tau)}_{t_1}\phi^{(s)}_{t_2}  -\phi^{(s)}^2_{t_1} \e^{-(\rmi\omega_0+\Gamma_0)(t_2-t_1)}\right]+
%\theta(t_1-t_2) \phi^{(0)}_{t_1}\phi^{(s)}_{t_2} \\
\phi^{(0)}_{t_1}\phi^{(s)}_{t_2} %\\\nonumber
+ \theta_{t_2-t_1}\phi^{(s)}_{t_1}\left[\phi^{(s)}_{t_2}  -\phi^{(s)}_{t_1} \e^{-(\rmi\omega_0+\Gamma_0) (t_2-t_1)}\right],
\end{align}
$t_< = {\rm min}(t_1,t_2)$, $\theta_t$ is the Heaviside step function, $\phi^{(\tau)}_t = \phi^{(0)}_t + \phi^{(s)}_t$ describes transmission of the single-photon pulse, and
$
 \phi^{(s)}_t = -\Gamma_0 \int_{-\infty}^t \phi^{(0)}_{t'} \e^{-(\rmi\omega_0+\Gamma_0)(t-t')}dt' \,.
$

As an example, we consider the incident pulse that has a Gaussian shape, 
\begin{align}\label{Eq:Gauss}
\phi^{(0)}_{t} = \frac{\sqrt{\gamma}}{\pi^{1/4}}\e^{-\rmi\omega t - \gamma^2 t^2/2} \,
\end{align}
with $\omega=\omega_0$ and $\gamma=0.5\Gamma_0$, that is shown in Fig.~\ref{fig:psi2}(a). 
The transmitted two-photon wave functions Eq.~\eqref{Eq:psi2} are plotted by color in Fig.~\ref{fig:psi2}(b,c). The probability of scattering in the state with two $X$-polarized photons, $|\psi_{XXx}(t_1,t_2)|^2$ [Fig.~\ref{fig:psi2}(b)], is significant only in the vicinity of the diagonal, $|t_1-t_2| \lesssim 1/\Gamma_0$. Indeed, if the two photons scatter at larger delays, they do so independently and each process is described by the transmission coefficients Eqs.~\eqref{Eq:st}~\cite{Rosenblum2015}. Note that at the central frequency of the pulse $t(\omega_0)=0$ and $s(\omega_0)=1$. Therefore, with dominant probability, the first photon is scattered in the $Y$ polarization and the atom switches to the $|y\rangle$ state~\cite{Pinotsi2008}. Then, the second photon passes the atom without interaction. Such case is described by $|\psi_{XYy}(t_1,t_2)|^2$ [Fig.~\ref{fig:psi2}(c)]. Its value is significant in the large region under the diagonal, $ t_1-t_2 \gtrsim 1/\Gamma_0$, which corresponds exactly to the described order of the scattering events.

%The probability of detecting an $X$-polarized photon at time $t_1$ and a $Y$-polarized photon at time $t_2$, $|\psi_{XYy}(t_1,t_2)|^2$ [Fig.~\ref{fig:psi2}(c)], 
%%%%%%%%%%%%%%%%%%%%%%%%%%
\begin{figure}[t]
\centering
\includegraphics[width=0.99\columnwidth]{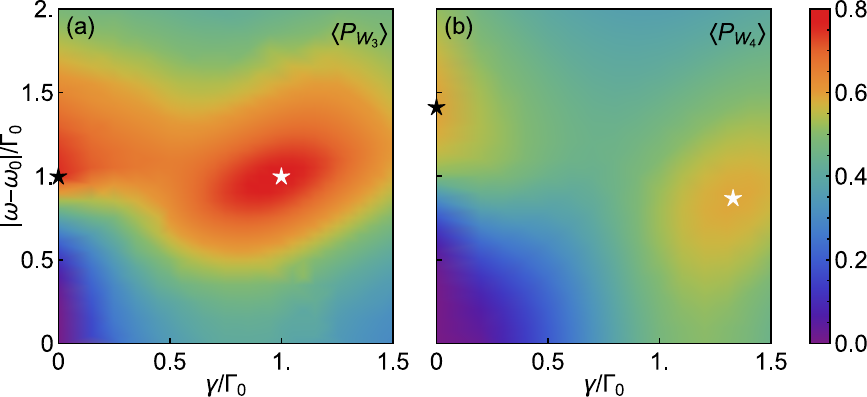}
\caption{(a) The expected value $\langle P_{W_3} \rangle $ of the probability of converting the final state of the atom and two photons into the canonical $W$ state by SLOCC. The calculation is performed after Eq.~\eqref{Eq:Pa} for the incident two-photon pulse with Gaussian envelope, Eq.~\eqref{Eq:Gauss}.   (b) $\langle P_{W_4} \rangle $ calculated for the three-photon pulse. Stars indicate positions of the maxima. 
}\label{fig:pw} 
\end{figure}
%%%%%%%%%%%%%%%%%%%%%%%%%%
%

{\it Tripartite entanglement.}
The wave function Eq.~\eqref{Eq:psi2} describes the the polarization state of the system provided the two transmitted photons were detected at the times $t_1$ and $t_2$. According to the classification of tree-qubit states based on stochastic local operations and classical communication (SLOCC)~\cite{Dur2000}, such state belongs to the $W$ class of the tripartite entanglement. Indeed, if we rename the photon polarization states $|X(Y)\rangle$ as $|0(1)\rangle$ and the atom states $|x(y)\rangle$ as $|1(0)\rangle$, the state Eq.~\eqref{Eq:psi2} turns to be the linear combination of the states $|001\rangle$, $|010\rangle$, and $|100\rangle$, which is precisely the generalized $W$ state. 

There is no conventional measure of tripartite entanglement~\cite{Ghne2009,Horodecki2009}. To quantify the entanglement of our state Eq.~\eqref{Eq:psi2}, we use the probability $P_{W_3}$ with which it can be converted to the canonical $W$ state  $(|001\rangle+|010\rangle+|100\rangle)/\sqrt{3}$ by SLOCC using the procedure described in Ref.~\cite{Yang2004}.  The probability is readily expressed via the coefficients of the wave function: 
\begin{align}\label{Eq:Pw}
    &P_{W_3} (t_1,t_2)=
    \frac{3\, {\rm min}[|\psi^{XXx}_{t_1,t_2}|^2, |\psi^{XYy}_{t_1,t_2}|^2, |\psi^{XYy}_{t_2,t_1}|^2]}{|\psi^{XXx}_{t_1,t_2}|^2 +|\psi^{XYy}_{t_1,t_2}|^2 +|\psi^{XYy}_{t_2,t_1}|^2}
    . 
\end{align}
The color plot of  $P_{W_3} (t_1,t_2)$ is shown in Fig,~\ref{fig:psi2}(d). The maximal values are  achieved near the diagonal where all thee coefficient in the wave function Eq.~\eqref{Eq:psiout} are of the same order. 

While $P_{W_3} (t_1,t_2)$ quantifies the entanglement of the atom and the photons detected at times $t_1$ and $t_2$, the total entanglement degree of the final state can be charecterized by the expected value of $P_{W_3} (t_1,t_2)$ obtained by averaging over $t_1$ and $t_2$, 
\begin{align}\label{Eq:Pa}
    &\langle P_{W_3} \rangle = 
    \frac{3\iint {\rm min}[|\psi^{XXx}_{t_1,t_2}|^2, |\psi^{XYy}_{t_1,t_2}|^2, |\psi^{XYy}_{t_2,t_1}|^2] dt_1dt_2 }{\iint [|\psi^{XXx}_{t_1,t_2}|^2+2 |\psi^{XYy}_{t_1,t_2}|^2] dt_1dt_2} \,.
\end{align}
Note that the denominator of Eq.~\eqref{Eq:Pa} is the norm of the final state and equals unity provided the incident pulse is normalized. 
For the parameters of Fig.~\ref{fig:psi2}, the overlap between $\psi^{XXx}$ and  $\psi^{XYy}$ is rather small, leading to small value of $\langle P_{W_3} \rangle$. To maximize $\langle P_{W_3}\rangle$, we tune the parameters of the incident Gaussian pulse Eq.~\eqref{Eq:Gauss}.
Figure~\ref{fig:pw}(a) shows the color plot of $\langle P_{W_3} \rangle$ as a function of the pulse central frequency $\omega$ and the spectral width $\gamma$. The dependence is symmetric with respect to $\omega=\omega_0$ and has two maxima. The one at $|\omega-\omega_0|= \Gamma_0$, $\gamma \to 0$ (black star) corresponds to the limit of a monochromatic wave. Then, straightforward analytical calculation yields
\begin{align}\label{Eq:Pw3}
\langle P_{W_3}\rangle_\omega= 3 |t(\omega)|^2\,{\rm min}\{|t(\omega)|^2, |s(\omega)|^2\}
\end{align}
that has the maximum value $\langle P_{W_3}\rangle_{\omega_0 \pm \Gamma_0} = 0.75$. A slightly larger value $\langle P_{W_3} \rangle \approx 0.77$ is achieved in the second maximum at $|\omega-\omega_0| \approx  0.98 \Gamma_0$, $\gamma \approx 0.97 \Gamma_0$ (white star), which corresponds to a short pulse. 
%Apparently, such parameters maximize the overlap between $\psi^{XXx}_{t_1,t_2}$ and $\psi^{XYy}_{t_1,t_2}$, see Supplement for the corresponding plots. 

%%%%%%%%%%%%%%%%%%%%%%%%%%
\begin{figure}[t]
\centering
\includegraphics[width=0.99\columnwidth]{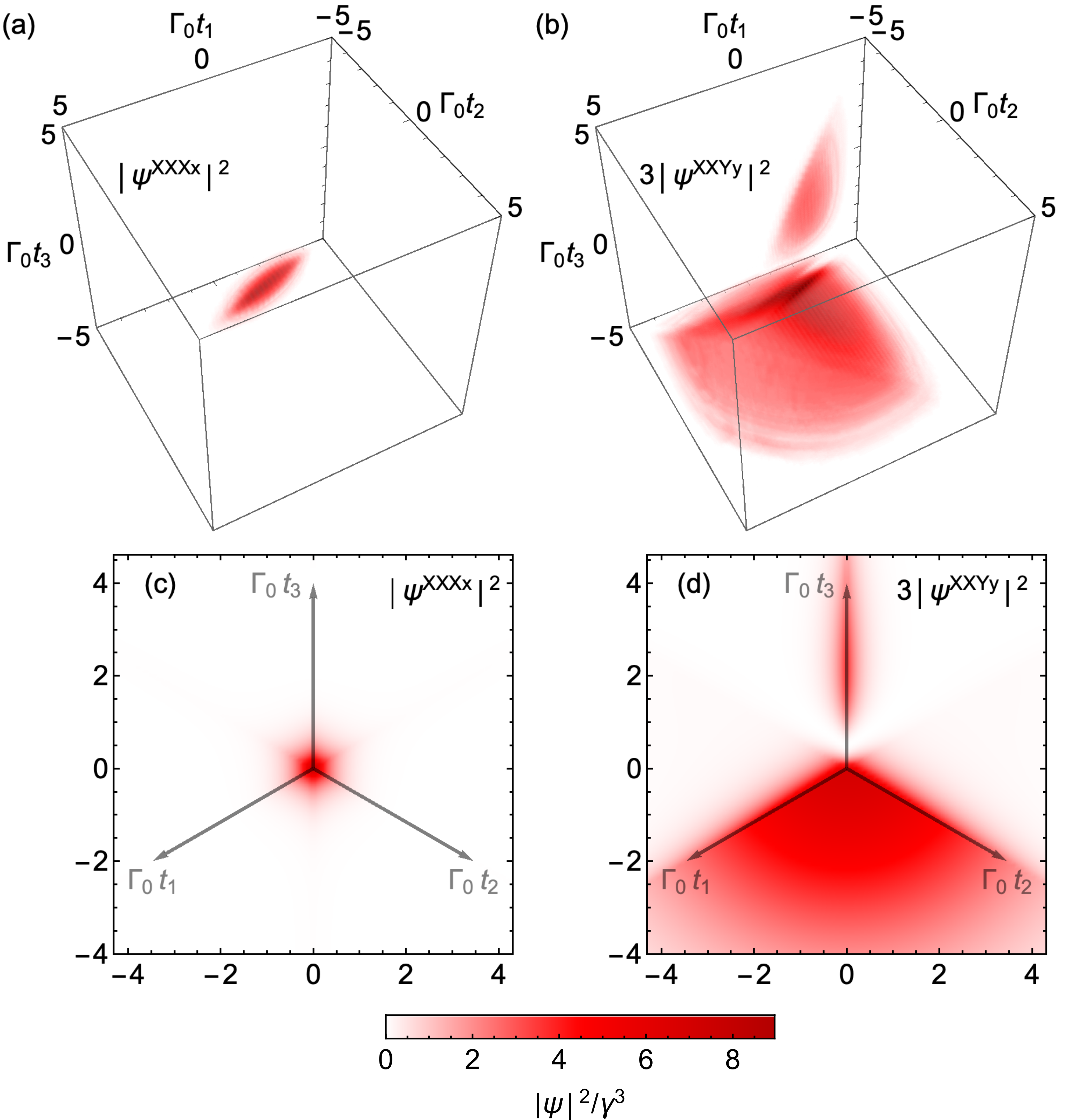}
\caption{(a,b) Volume plot of the real-time wave functions of the final state for the three-photon scattering process. (c,d) Cross-section of the plots in (a,b) in the direction perpendicular to the main diagonal at $t_1+t_2+t_3=0$. Calculation is performed after Eqs.~\eqref{Eq:XXXx}--\eqref{Eq:XXYy} for the incident three-photon pulse with Gaussian envelope, Eq.~\eqref{Eq:Gauss} with $\omega=\omega_0$, $\gamma=0.2\Gamma_0$. 
}\label{fig:psi3} 
\end{figure}
%%%%%%%%%%%%%%%%%%%%%%%%%%
%

{\it Three-photon scattering}.
The above results are easily generalized for the larger number of incident photons. To show this, we consider three-photon scattering. The diagram corresponding to the non-trivial part of the process is shown in Fig.~\ref{fig:dia}~\footnote{See Supplementary for the explicit expressions for the scattering matrix elements}. When there are three identical $X$-polarized photons in incident state, $\psi^{\rm (in)}_{t_1,t_2,t_3} = \phi^{(0)}_{t_1}\phi^{(0)}_{t_2}\phi^{(0)}_{t_3} |XXXx\rangle$, the final state reads
%
%\begin{widetext}
%\begin{align} \label{Eq:XXXx}
%\psi^{XXXx}_{t_1,t_2,t_3}
%&= \phi^{(\tau)}_{t_1}\phi^{(\tau)}_{t_2}\phi^{(\tau)}_{t_3} + \phi^{(s)}^2_{t_(1)}[\phi^{(s)}_{t_{(2)}}-\phi^{(0)}_{t_{(2)}}]
%\e^{-(\rmi\omega_0+\Gamma_0)(t_{(3)}-t_{(1)})} \\\nonumber
%&- \phi^{(s)}^2_{t_(1)}\phi^{(\tau)}_{t_{(3)}}
%\e^{-(\rmi\omega_0+\Gamma_0)(t_{(2)}-t_{(1)})} 
%- \phi^{(s)}^2_{t_{(2)}}\phi^{(\tau)}_{t_(1)}
%\e^{-(\rmi\omega_0+\Gamma_0)(t_{(3)}-t_{(2)})} 
% \,,
%\\
%\psi^{XXYy}_{t_1,t_2,t_3}
%&=  \label{Eq:XXYy} 
%\theta(t_>-t_3)\theta(t_3-t_<)\phi^{(0)}_{t_>}\big\{  \phi^{(\tau)}_{t_<}\phi^{(s)}_{t_3} 
%-\phi^{(s)}^2_{t_<} \e^{-(\rmi\omega_0+\Gamma_0)(t_3-t_<)} 
%\big\} \\\nonumber
%&+\theta(t_3-t_>)\big\{  \phi^{(\tau)}_{t_<}\phi^{(\tau)}_{t_>}\phi^{(s)}_{t_3} 
%-\phi^{(s)}^2_{t_<}\phi^{(s)}_{t_3} e^{-(\rmi\omega_0+\Gamma_0)(t_>-t_<)} 
%-\phi^{(s)}^2_{t_>}\phi^{(\tau)}_{t_<} e^{-(\rmi\omega_0+\Gamma_0)(t_3-t_>)} \\\nonumber
%&+\phi^{(s)}^2_{t_<}[\phi^{(s)}_{t_>}-\phi^{(0)}_{t_>}] e^{-(\rmi\omega_0+\Gamma_0)(t_3-t_<)} 
%\big\} 
%+\theta(t_<-t_3)  \phi^{(0)}_{t_<}\phi^{(0)}_{t_>}\phi^{(s)}_{t_3} \,,
%\end{align}
%\end{widetext}
%
%
%
\begin{align}\label{Eq:psi4}
\psi^{\rm (out)}_{t_1,t_2,t_3}& = 
\psi^{XXXx}_{t_1,t_2,t_3} |XXXx\rangle +
\psi^{XXYy}_{t_1,t_2,t_3} |XXYy\rangle \nonumber\\
&+
\psi^{XXYy}_{t_1,t_3,t_2} |XYXy\rangle +
\psi^{XXYy}_{t_3,t_1,t_2} |YXXy\rangle ,
\end{align}
where
\begin{align} 
&\psi^{XXXx}_{t_1,t_2,t_3}
= \phi^{(\tau)}_{t_1}\phi^{(\tau)}_{t_2}\phi^{(\tau)}_{t_3} \nonumber\\\label{Eq:XXXx}
&\hspace{2cm}- [\phi^{(s)}_{t_{(1)}}]^2\phi^{(\tau)}_{t_{(3)}}
\e^{-(\rmi\omega_0+\Gamma_0)(t_{(2)}-t_{(1)})} \\\nonumber
&\hspace{2cm}- [\phi^{(s)}_{t_{(2)}}]^2\phi^{(\tau)}_{t_{(1)}}
\e^{-(\rmi\omega_0+\Gamma_0)(t_{(3)}-t_{(2)})} \\\nonumber
&\hspace{2cm}+ [\phi^{(s)}_{t_{(1)}}]^2[\phi^{(s)}_{t_{(2)}}-\phi^{(0)}_{t_{(2)}}]
\e^{-(\rmi\omega_0+\Gamma_0)(t_{(3)}-t_{(1)})} \,, \\
&\psi^{XXYy}_{t_1,t_2,t_3}
=\theta_{t_<-t_3} \phi^{(0)}_{t_<}\phi^{(0)}_{t_>}\phi^{(s)}_{t_3}   
+ \theta_{t_3-t_>}\big\{  \phi^{(\tau)}_{t_<}\phi^{(\tau)}_{t_>}\phi^{(s)}_{t_3} \nonumber\\
&\hspace{2cm}-[\phi^{(s)}_{t_<}]^2\phi^{(s)}_{t_3} \e^{-(\rmi\omega_0+\Gamma_0)(t_>-t_<)} \nonumber\\
&\hspace{2cm}-[\phi^{(s)}_{t_>}]^2\phi^{(\tau)}_{t_<} \e^{-(\rmi\omega_0+\Gamma_0)(t_3-t_>)} \label{Eq:XXYy}  \\\nonumber
&\hspace{2cm}+[\phi^{(s)}_{t_<}]^2[\phi^{(s)}_{t_>}-\phi^{(0)}_{t_>}] \e^{-(\rmi\omega_0+\Gamma_0)(t_3-t_<)} 
\big\}
\\ \nonumber
&+\theta_{t_>-t_3}\theta_{t_3-t_<}\phi^{(0)}_{t_>}\big\{  \phi^{(\tau)}_{t_<}\phi^{(s)}_{t_3} 
-[\phi^{(s)}_{t_<}]^2 \e^{-(\rmi\omega_0+\Gamma_0)(t_3-t_<)} 
\big\},
\end{align}
$t_{(1)} \leq t_{(2)} \leq t_{(3)}$ are the times $t_1$, $t_2$, $t_3$ sorted in the ascending order, $t_< = {\rm min}(t_1,t_2)$ and  $t_>= {\rm max}(t_1,t_2)$. 

Figure~\ref{fig:psi3} shows $|\psi^{XXXx}_{t_1,t_2,t_3}|^2$ and $|\psi^{XXYy}_{t_1,t_2,t_3}|^2$ calculated for the Gaussian incident pulse. The probability to detect all the transmitted photons in the $X$ polarization is nonzero only if all three detection times are close, $|t_1-t_2|,|t_2-t_3|,|t_3-t_1| \lesssim 1/\Gamma_0$. This corresponds to the main diagonal of the coordinate frame in Fig.~\ref{fig:psi3}(a) and the center of the cross-section shown in Fig.~\ref{fig:psi3}(c). Similarly to the two-photon case, that is explained by the fact that if at least one of the photons is detected with a larger delay, that means that it has scattered independently and must convert to $Y$ polarization, since $s(\omega_0)=1$. 
The probability that one of the photons is detected in $Y$ polarization at time $t_3$ is shown in Figs.~\ref{fig:psi3}(b,d) and has a much more peculiar distribution. 
If the three photons are detected with large delay, it is the first of them that will be converted to $Y$ polarization, see the region in the bottom of the cross-section Fig.~\ref{fig:psi3}(d) which corresponds to $t_3 \lesssim t_1,t_2$. Alternatively, it could happen that the first two photons are transmitted simultaneously so that they keep their $X$ polarization, which corresponds to the two-photon scattering amplitude $|\psi^{XXx}_{t_1,t_2}|^2$ depicted in Fig.~\ref{fig:psi2}(b). Then, the last of the three photons will be converted to $Y$ polarization, see the line in the upper part of the cross-section Figs.~\ref{fig:psi3}(d) which corresponds to $t_3 \gtrsim t_1 \approx t_2$.  

The entanglement of the four-partite wave function describing the final state of the atom and three photons Eq.~\eqref{Eq:psi4} can be quantified by the averaged probability $\langle P_{W_4}\rangle$  that is defined in a manner similar to the three-partite case~\cite{Yang2004}. Dependence of $\langle P_{W_4}\rangle$ on the parameters of the incident Gaussian pulse are shown in Fig.~\ref{fig:pw}(b). As in the case of two photons, it has two maxima corresponding to monochromatic and short incident pulses. One that corresponds to monochromatic limit is at $|\omega - \omega_0 | = \sqrt{2} \Gamma_0$, $\gamma \to 0$  (black star) and gives $\langle P_{W_4}\rangle = 16/27 \approx 0.59$~\footnote{For monochromatic three-photon incident pulse calculation yields $\langle P_{W_4}\rangle_\omega= 4 |t(\omega)|^4\,{\rm min}\{|t(\omega)|^2, |s(\omega)|^2\}$}, the other is at $|\omega -\omega_0| \approx  0.87\Gamma_0$, $\gamma \approx 1.33\Gamma_0$ (white star) and gives a close value $\langle P_{W_4}\rangle  \approx 0.59$.

To further improve the values of $\langle P_{W_3}\rangle$ and $\langle P_{W_4}\rangle$, the pulse shape should be optimized. By adding to the pulse higher temporal modes described by Hermite polynomials we were able to achieve $\langle P_{W_3}\rangle  \approx 0.8$ and $\langle P_{W_4}\rangle  \approx 0.62$, see Supplementary for details. 
%To maximize it, the overlap between the $|\psi^{XXXx}_{t_1,t_2,t_3}|^2$ and $|\psi^{XXYy}_{t_1,t_2,t_3}|^2$ should be optimized. For Gaussian pulse the maximum $\langle P_{W_3}\rangle = ?$ is achieved at ???

%From the unitarity of the scattering matrix, it follows that
%\begin{align}
%\iiint \left(|\psi_{XXX}(t_1,t_2,t_3)|^2 + 3 |\psi_{XXY}(t_1,t_2,t_3)|^2 \right|) \,dt_1 dt_2 dt_3 = \left( \int |\phi^{(0)}(t)|^2 dt \right)^2
%\end{align}
%which we have also checked numerically.

{\it Outlook.}
We have proposed a scheme for single-shot generation of multipartite polarization-entangled $W$ states of a $\Lambda$-type atom and several photons in a waveguide. Given the certain robustness of $W$-state, the purely photonic $W$ states can be obtained simply by disregarding the atom. We note that if the two optical transitions of the $\Lambda$ atom differ in the photon propagation direction or frequency~\cite{Bradford2012}, then the frequency- or direction-entangled photons can be generated. While we presented the theory for the case of chiral coupling, the photon transmission in the non-chiral case is described by exactly the same equations but with the twice smaller values of  $s(\omega)$ and $\varphi(s)$. 
Another possible generalization is to consider a system of several $\Lambda$ atoms excited by several photons in a waveguide. Qualitative analysis shows that that would enable generation of an entangled  $W$ state of all the atoms and the photons. However, the calculation of such scattering process in a non-chiral setup is far not straightforward and shall be the subject of future research. 

{\it Acknowledgements.} We are grateful to A.N. Poddubny, I.V. Iorsh, and D.S. Smirnov for useful discussions.

\bibliography{lambda.bib}

\begin{thebibliography}{48}%
\makeatletter
\providecommand \@ifxundefined [1]{%
 \@ifx{#1\undefined}
}%
\providecommand \@ifnum [1]{%
 \ifnum #1\expandafter \@firstoftwo
 \else \expandafter \@secondoftwo
 \fi
}%
\providecommand \@ifx [1]{%
 \ifx #1\expandafter \@firstoftwo
 \else \expandafter \@secondoftwo
 \fi
}%
\providecommand \natexlab [1]{#1}%
\providecommand \enquote  [1]{``#1''}%
\providecommand \bibnamefont  [1]{#1}%
\providecommand \bibfnamefont [1]{#1}%
\providecommand \citenamefont [1]{#1}%
\providecommand \href@noop [0]{\@secondoftwo}%
\providecommand \href [0]{\begingroup \@sanitize@url \@href}%
\providecommand \@href[1]{\@@startlink{#1}\@@href}%
\providecommand \@@href[1]{\endgroup#1\@@endlink}%
\providecommand \@sanitize@url [0]{\catcode `\\12\catcode `\$12\catcode
  `\&12\catcode `\#12\catcode `\^12\catcode `\_12\catcode `\%12\relax}%
\providecommand \@@startlink[1]{}%
\providecommand \@@endlink[0]{}%
\providecommand \url  [0]{\begingroup\@sanitize@url \@url }%
\providecommand \@url [1]{\endgroup\@href {#1}{\urlprefix }}%
\providecommand \urlprefix  [0]{URL }%
\providecommand \Eprint [0]{\href }%
\providecommand \doibase [0]{http://dx.doi.org/}%
\providecommand \selectlanguage [0]{\@gobble}%
\providecommand \bibinfo  [0]{\@secondoftwo}%
\providecommand \bibfield  [0]{\@secondoftwo}%
\providecommand \translation [1]{[#1]}%
\providecommand \BibitemOpen [0]{}%
\providecommand \bibitemStop [0]{}%
\providecommand \bibitemNoStop [0]{.\EOS\space}%
\providecommand \EOS [0]{\spacefactor3000\relax}%
\providecommand \BibitemShut  [1]{\csname bibitem#1\endcsname}%
\let\auto@bib@innerbib\@empty
%</preamble>
\bibitem [{\citenamefont {G\"{u}hne}\ and\ \citenamefont
  {T{\'{o}}th}(2009)}]{Ghne2009}%
  \BibitemOpen
  \bibfield  {author} {\bibinfo {author} {\bibfnamefont {O.}~\bibnamefont
  {G\"{u}hne}}\ and\ \bibinfo {author} {\bibfnamefont {G.}~\bibnamefont
  {T{\'{o}}th}},\ }\bibfield  {title} {\enquote {\bibinfo {title} {Entanglement
  detection},}\ }\href {\doibase 10.1016/j.physrep.2009.02.004} {\bibfield
  {journal} {\bibinfo  {journal} {Physics Reports}\ }\textbf {\bibinfo {volume}
  {474}},\ \bibinfo {pages} {1} (\bibinfo {year} {2009})}\BibitemShut {NoStop}%
\bibitem [{\citenamefont {Horodecki}\ \emph {et~al.}(2009)\citenamefont
  {Horodecki}, \citenamefont {Horodecki}, \citenamefont {Horodecki},\ and\
  \citenamefont {Horodecki}}]{Horodecki2009}%
  \BibitemOpen
  \bibfield  {author} {\bibinfo {author} {\bibfnamefont {R.}~\bibnamefont
  {Horodecki}}, \bibinfo {author} {\bibfnamefont {P.}~\bibnamefont
  {Horodecki}}, \bibinfo {author} {\bibfnamefont {M.}~\bibnamefont
  {Horodecki}}, \ and\ \bibinfo {author} {\bibfnamefont {K.}~\bibnamefont
  {Horodecki}},\ }\bibfield  {title} {\enquote {\bibinfo {title} {Quantum
  entanglement},}\ }\href {\doibase 10.1103/revmodphys.81.865} {\bibfield
  {journal} {\bibinfo  {journal} {Rev. Mod. Phys.}\ }\textbf {\bibinfo {volume}
  {81}},\ \bibinfo {pages} {865} (\bibinfo {year} {2009})}\BibitemShut
  {NoStop}%
\bibitem [{\citenamefont {Karlsson}\ and\ \citenamefont
  {Bourennane}(1998)}]{Karlsson1998}%
  \BibitemOpen
  \bibfield  {author} {\bibinfo {author} {\bibfnamefont {A.}~\bibnamefont
  {Karlsson}}\ and\ \bibinfo {author} {\bibfnamefont {M.}~\bibnamefont
  {Bourennane}},\ }\bibfield  {title} {\enquote {\bibinfo {title} {Quantum
  teleportation using three-particle entanglement},}\ }\href {\doibase
  10.1103/PhysRevA.58.4394} {\bibfield  {journal} {\bibinfo  {journal} {Phys.
  Rev. A}\ }\textbf {\bibinfo {volume} {58}},\ \bibinfo {pages} {4394--4400}
  (\bibinfo {year} {1998})}\BibitemShut {NoStop}%
\bibitem [{\citenamefont {D\"ur}\ \emph {et~al.}(2000)\citenamefont {D\"ur},
  \citenamefont {Vidal},\ and\ \citenamefont {Cirac}}]{Dur2000}%
  \BibitemOpen
  \bibfield  {author} {\bibinfo {author} {\bibfnamefont {W.}~\bibnamefont
  {D\"ur}}, \bibinfo {author} {\bibfnamefont {G.}~\bibnamefont {Vidal}}, \ and\
  \bibinfo {author} {\bibfnamefont {J.~I.}\ \bibnamefont {Cirac}},\ }\bibfield
  {title} {\enquote {\bibinfo {title} {Three qubits can be entangled in two
  inequivalent ways},}\ }\href {\doibase 10.1103/PhysRevA.62.062314} {\bibfield
   {journal} {\bibinfo  {journal} {Phys. Rev. A}\ }\textbf {\bibinfo {volume}
  {62}},\ \bibinfo {pages} {062314} (\bibinfo {year} {2000})}\BibitemShut
  {NoStop}%
\bibitem [{\citenamefont {Agrawal}\ and\ \citenamefont
  {Pati}(2006)}]{Agrawal2006}%
  \BibitemOpen
  \bibfield  {author} {\bibinfo {author} {\bibfnamefont {P.}~\bibnamefont
  {Agrawal}}\ and\ \bibinfo {author} {\bibfnamefont {A.}~\bibnamefont {Pati}},\
  }\bibfield  {title} {\enquote {\bibinfo {title} {Perfect teleportation and
  superdense coding with {W} states},}\ }\href {\doibase
  10.1103/PhysRevA.74.062320} {\bibfield  {journal} {\bibinfo  {journal} {Phys.
  Rev. A}\ }\textbf {\bibinfo {volume} {74}},\ \bibinfo {pages} {062320}
  (\bibinfo {year} {2006})}\BibitemShut {NoStop}%
\bibitem [{\citenamefont {Briegel}\ and\ \citenamefont
  {Raussendorf}(2001)}]{Briegel2001}%
  \BibitemOpen
  \bibfield  {author} {\bibinfo {author} {\bibfnamefont {H.~J.}\ \bibnamefont
  {Briegel}}\ and\ \bibinfo {author} {\bibfnamefont {R.}~\bibnamefont
  {Raussendorf}},\ }\bibfield  {title} {\enquote {\bibinfo {title} {Persistent
  entanglement in arrays of interacting particles},}\ }\href {\doibase
  10.1103/PhysRevLett.86.910} {\bibfield  {journal} {\bibinfo  {journal} {Phys.
  Rev. Lett.}\ }\textbf {\bibinfo {volume} {86}},\ \bibinfo {pages} {910--913}
  (\bibinfo {year} {2001})}\BibitemShut {NoStop}%
\bibitem [{\citenamefont {Raussendorf}\ and\ \citenamefont
  {Briegel}(2001)}]{Raussendorf2001}%
  \BibitemOpen
  \bibfield  {author} {\bibinfo {author} {\bibfnamefont {R.}~\bibnamefont
  {Raussendorf}}\ and\ \bibinfo {author} {\bibfnamefont {H.~J.}\ \bibnamefont
  {Briegel}},\ }\bibfield  {title} {\enquote {\bibinfo {title} {A one-way
  quantum computer},}\ }\href {\doibase 10.1103/PhysRevLett.86.5188} {\bibfield
   {journal} {\bibinfo  {journal} {Phys. Rev. Lett.}\ }\textbf {\bibinfo
  {volume} {86}},\ \bibinfo {pages} {5188} (\bibinfo {year}
  {2001})}\BibitemShut {NoStop}%
\bibitem [{\citenamefont {Knill}\ \emph {et~al.}(2001)\citenamefont {Knill},
  \citenamefont {Laflamme},\ and\ \citenamefont {Milburn}}]{Knill2001}%
  \BibitemOpen
  \bibfield  {author} {\bibinfo {author} {\bibfnamefont {E.}~\bibnamefont
  {Knill}}, \bibinfo {author} {\bibfnamefont {R.}~\bibnamefont {Laflamme}}, \
  and\ \bibinfo {author} {\bibfnamefont {G.~J.}\ \bibnamefont {Milburn}},\
  }\bibfield  {title} {\enquote {\bibinfo {title} {A scheme for efficient
  quantum computation with linear optics},}\ }\href {\doibase 10.1038/35051009}
  {\bibfield  {journal} {\bibinfo  {journal} {Nature}\ }\textbf {\bibinfo
  {volume} {409}},\ \bibinfo {pages} {46} (\bibinfo {year} {2001})}\BibitemShut
  {NoStop}%
\bibitem [{\citenamefont {R.}\ \emph {et~al.}(2017)\citenamefont {R.},
  \citenamefont {Wilson},\ and\ \citenamefont {Firstenberg}}]{Roy2017}%
  \BibitemOpen
  \bibfield  {author} {\bibinfo {author} {\bibfnamefont {Dibyendu}\
  \bibnamefont {R.}}, \bibinfo {author} {\bibfnamefont {C.~M.}\ \bibnamefont
  {Wilson}}, \ and\ \bibinfo {author} {\bibfnamefont {O.}~\bibnamefont
  {Firstenberg}},\ }\bibfield  {title} {\enquote {\bibinfo {title} {Colloquium:
  Strongly interacting photons in one-dimensional continuum},}\ }\href
  {\doibase 10.1103/RevModPhys.89.021001} {\bibfield  {journal} {\bibinfo
  {journal} {Rev. Mod. Phys.}\ }\textbf {\bibinfo {volume} {89}},\ \bibinfo
  {pages} {021001} (\bibinfo {year} {2017})}\BibitemShut {NoStop}%
\bibitem [{\citenamefont {Chang}\ \emph {et~al.}(2018)\citenamefont {Chang},
  \citenamefont {Douglas}, \citenamefont {Gonz\'alez-Tudela}, \citenamefont
  {Hung},\ and\ \citenamefont {Kimble}}]{Chang2018}%
  \BibitemOpen
  \bibfield  {author} {\bibinfo {author} {\bibfnamefont {D.~E.}\ \bibnamefont
  {Chang}}, \bibinfo {author} {\bibfnamefont {J.~S.}\ \bibnamefont {Douglas}},
  \bibinfo {author} {\bibfnamefont {A.}~\bibnamefont {Gonz\'alez-Tudela}},
  \bibinfo {author} {\bibfnamefont {C.-L.}\ \bibnamefont {Hung}}, \ and\
  \bibinfo {author} {\bibfnamefont {H.~J.}\ \bibnamefont {Kimble}},\ }\bibfield
   {title} {\enquote {\bibinfo {title} {Colloquium: Quantum matter built from
  nanoscopic lattices of atoms and photons},}\ }\href {\doibase
  10.1103/RevModPhys.90.031002} {\bibfield  {journal} {\bibinfo  {journal}
  {Rev. Mod. Phys.}\ }\textbf {\bibinfo {volume} {90}},\ \bibinfo {pages}
  {031002} (\bibinfo {year} {2018})}\BibitemShut {NoStop}%
\bibitem [{\citenamefont {Sheremet}\ \emph {et~al.}(2023)\citenamefont
  {Sheremet}, \citenamefont {Petrov}, \citenamefont {Iorsh}, \citenamefont
  {Poshakinskiy},\ and\ \citenamefont {Poddubny}}]{Sheremet2023}%
  \BibitemOpen
  \bibfield  {author} {\bibinfo {author} {\bibfnamefont {A.~S.}\ \bibnamefont
  {Sheremet}}, \bibinfo {author} {\bibfnamefont {M.~I.}\ \bibnamefont
  {Petrov}}, \bibinfo {author} {\bibfnamefont {I.~V.}\ \bibnamefont {Iorsh}},
  \bibinfo {author} {\bibfnamefont {A.~V.}\ \bibnamefont {Poshakinskiy}}, \
  and\ \bibinfo {author} {\bibfnamefont {A.~N.}\ \bibnamefont {Poddubny}},\
  }\bibfield  {title} {\enquote {\bibinfo {title} {Waveguide quantum
  electrodynamics: Collective radiance and photon-photon correlations},}\
  }\href {\doibase 10.1103/RevModPhys.95.015002} {\bibfield  {journal}
  {\bibinfo  {journal} {Rev. Mod. Phys.}\ }\textbf {\bibinfo {volume} {95}},\
  \bibinfo {pages} {015002} (\bibinfo {year} {2023})}\BibitemShut {NoStop}%
\bibitem [{\citenamefont {Ralph}\ \emph {et~al.}(2015)\citenamefont {Ralph},
  \citenamefont {S\"ollner}, \citenamefont {Mahmoodian}, \citenamefont
  {White},\ and\ \citenamefont {Lodahl}}]{Ralph2015}%
  \BibitemOpen
  \bibfield  {author} {\bibinfo {author} {\bibfnamefont {T.~C.}\ \bibnamefont
  {Ralph}}, \bibinfo {author} {\bibfnamefont {I.}~\bibnamefont {S\"ollner}},
  \bibinfo {author} {\bibfnamefont {S.}~\bibnamefont {Mahmoodian}}, \bibinfo
  {author} {\bibfnamefont {A.~G.}\ \bibnamefont {White}}, \ and\ \bibinfo
  {author} {\bibfnamefont {P.}~\bibnamefont {Lodahl}},\ }\bibfield  {title}
  {\enquote {\bibinfo {title} {Photon sorting, efficient bell measurements, and
  a deterministic controlled-$z$ gate using a passive two-level
  nonlinearity},}\ }\href {\doibase 10.1103/PhysRevLett.114.173603} {\bibfield
  {journal} {\bibinfo  {journal} {Phys. Rev. Lett.}\ }\textbf {\bibinfo
  {volume} {114}},\ \bibinfo {pages} {173603} (\bibinfo {year}
  {2015})}\BibitemShut {NoStop}%
\bibitem [{\citenamefont {Ilin}\ \emph {et~al.}(2023)\citenamefont {Ilin},
  \citenamefont {Poshakinskiy}, \citenamefont {Poddubny},\ and\ \citenamefont
  {Iorsh}}]{Ilin2023}%
  \BibitemOpen
  \bibfield  {author} {\bibinfo {author} {\bibfnamefont {D.}~\bibnamefont
  {Ilin}}, \bibinfo {author} {\bibfnamefont {A.~V.}\ \bibnamefont
  {Poshakinskiy}}, \bibinfo {author} {\bibfnamefont {A.~N.}\ \bibnamefont
  {Poddubny}}, \ and\ \bibinfo {author} {\bibfnamefont {I.}~\bibnamefont
  {Iorsh}},\ }\bibfield  {title} {\enquote {\bibinfo {title} {Frequency combs
  with parity-protected cross-correlations and entanglement from dynamically
  modulated qubit arrays},}\ }\href {\doibase 10.1103/PhysRevLett.130.023601}
  {\bibfield  {journal} {\bibinfo  {journal} {Phys. Rev. Lett.}\ }\textbf
  {\bibinfo {volume} {130}},\ \bibinfo {pages} {023601} (\bibinfo {year}
  {2023})}\BibitemShut {NoStop}%
\bibitem [{\citenamefont {Pinotsi}\ and\ \citenamefont
  {Imamoglu}(2008)}]{Pinotsi2008}%
  \BibitemOpen
  \bibfield  {author} {\bibinfo {author} {\bibfnamefont {D.}~\bibnamefont
  {Pinotsi}}\ and\ \bibinfo {author} {\bibfnamefont {A.}~\bibnamefont
  {Imamoglu}},\ }\bibfield  {title} {\enquote {\bibinfo {title} {Single photon
  absorption by a single quantum emitter},}\ }\href {\doibase
  10.1103/PhysRevLett.100.093603} {\bibfield  {journal} {\bibinfo  {journal}
  {Phys. Rev. Lett.}\ }\textbf {\bibinfo {volume} {100}},\ \bibinfo {pages}
  {093603} (\bibinfo {year} {2008})}\BibitemShut {NoStop}%
\bibitem [{\citenamefont {Rosenblum}\ \emph {et~al.}(2015)\citenamefont
  {Rosenblum}, \citenamefont {Bechler}, \citenamefont {Shomroni}, \citenamefont
  {Lovsky}, \citenamefont {Guendelman},\ and\ \citenamefont
  {Dayan}}]{Rosenblum2015}%
  \BibitemOpen
  \bibfield  {author} {\bibinfo {author} {\bibfnamefont {S.}~\bibnamefont
  {Rosenblum}}, \bibinfo {author} {\bibfnamefont {O.}~\bibnamefont {Bechler}},
  \bibinfo {author} {\bibfnamefont {I.}~\bibnamefont {Shomroni}}, \bibinfo
  {author} {\bibfnamefont {Y.}~\bibnamefont {Lovsky}}, \bibinfo {author}
  {\bibfnamefont {G.}~\bibnamefont {Guendelman}}, \ and\ \bibinfo {author}
  {\bibfnamefont {B.}~\bibnamefont {Dayan}},\ }\bibfield  {title} {\enquote
  {\bibinfo {title} {Extraction of a single photon from an optical pulse},}\
  }\href {\doibase 10.1038/nphoton.2015.227} {\bibfield  {journal} {\bibinfo
  {journal} {Nat. Photonics}\ }\textbf {\bibinfo {volume} {10}},\ \bibinfo
  {pages} {19} (\bibinfo {year} {2015})}\BibitemShut {NoStop}%
\bibitem [{\citenamefont {Koshino}\ \emph {et~al.}(2010)\citenamefont
  {Koshino}, \citenamefont {Ishizaka},\ and\ \citenamefont
  {Nakamura}}]{Koshino2010}%
  \BibitemOpen
  \bibfield  {author} {\bibinfo {author} {\bibfnamefont {K.}~\bibnamefont
  {Koshino}}, \bibinfo {author} {\bibfnamefont {S.}~\bibnamefont {Ishizaka}}, \
  and\ \bibinfo {author} {\bibfnamefont {Y.}~\bibnamefont {Nakamura}},\
  }\bibfield  {title} {\enquote {\bibinfo {title} {Deterministic photon-photon
  $\sqrt{\text{{s}{w}{a}{p}}}$ gate using a $\ensuremath{\Lambda}$ system},}\
  }\href {\doibase 10.1103/PhysRevA.82.010301} {\bibfield  {journal} {\bibinfo
  {journal} {Phys. Rev. A}\ }\textbf {\bibinfo {volume} {82}},\ \bibinfo
  {pages} {010301} (\bibinfo {year} {2010})}\BibitemShut {NoStop}%
\bibitem [{\citenamefont {Rosenblum}\ \emph {et~al.}(2017)\citenamefont
  {Rosenblum}, \citenamefont {Borne},\ and\ \citenamefont
  {Dayan}}]{Rosenblum2017}%
  \BibitemOpen
  \bibfield  {author} {\bibinfo {author} {\bibfnamefont {S.}~\bibnamefont
  {Rosenblum}}, \bibinfo {author} {\bibfnamefont {A.}~\bibnamefont {Borne}}, \
  and\ \bibinfo {author} {\bibfnamefont {B.}~\bibnamefont {Dayan}},\ }\bibfield
   {title} {\enquote {\bibinfo {title} {Analysis of deterministic swapping of
  photonic and atomic states through single-photon raman interaction},}\ }\href
  {\doibase 10.1103/PhysRevA.95.033814} {\bibfield  {journal} {\bibinfo
  {journal} {Phys. Rev. A}\ }\textbf {\bibinfo {volume} {95}},\ \bibinfo
  {pages} {033814} (\bibinfo {year} {2017})}\BibitemShut {NoStop}%
\bibitem [{\citenamefont {Bechler}\ \emph {et~al.}(2018)\citenamefont
  {Bechler}, \citenamefont {Borne}, \citenamefont {Rosenblum}, \citenamefont
  {Guendelman}, \citenamefont {Mor}, \citenamefont {Netser}, \citenamefont
  {Ohana}, \citenamefont {Aqua}, \citenamefont {Drucker}, \citenamefont
  {Finkelstein}, \citenamefont {Lovsky}, \citenamefont {Bruch}, \citenamefont
  {Gurovich}, \citenamefont {Shafir},\ and\ \citenamefont
  {Dayan}}]{Bechler2018}%
  \BibitemOpen
  \bibfield  {author} {\bibinfo {author} {\bibfnamefont {O.}~\bibnamefont
  {Bechler}}, \bibinfo {author} {\bibfnamefont {A.}~\bibnamefont {Borne}},
  \bibinfo {author} {\bibfnamefont {S.}~\bibnamefont {Rosenblum}}, \bibinfo
  {author} {\bibfnamefont {G.}~\bibnamefont {Guendelman}}, \bibinfo {author}
  {\bibfnamefont {O.E.}\ \bibnamefont {Mor}}, \bibinfo {author} {\bibfnamefont
  {M.}~\bibnamefont {Netser}}, \bibinfo {author} {\bibfnamefont
  {T.}~\bibnamefont {Ohana}}, \bibinfo {author} {\bibfnamefont
  {Z.}~\bibnamefont {Aqua}}, \bibinfo {author} {\bibfnamefont {N.}~\bibnamefont
  {Drucker}}, \bibinfo {author} {\bibfnamefont {R.}~\bibnamefont
  {Finkelstein}}, \bibinfo {author} {\bibfnamefont {Y.}~\bibnamefont {Lovsky}},
  \bibinfo {author} {\bibfnamefont {R.}~\bibnamefont {Bruch}}, \bibinfo
  {author} {\bibfnamefont {D.}~\bibnamefont {Gurovich}}, \bibinfo {author}
  {\bibfnamefont {E.}~\bibnamefont {Shafir}}, \ and\ \bibinfo {author}
  {\bibfnamefont {B.}~\bibnamefont {Dayan}},\ }\bibfield  {title} {\enquote
  {\bibinfo {title} {A passive photon{\textendash}atom qubit swap operation},}\
  }\href {\doibase 10.1038/s41567-018-0241-6} {\bibfield  {journal} {\bibinfo
  {journal} {Nat. Physics}\ }\textbf {\bibinfo {volume} {14}},\ \bibinfo
  {pages} {996} (\bibinfo {year} {2018})}\BibitemShut {NoStop}%
\bibitem [{\citenamefont {Aqua}\ \emph {et~al.}(2019)\citenamefont {Aqua},
  \citenamefont {Kim},\ and\ \citenamefont {Dayan}}]{Aqua2019}%
  \BibitemOpen
  \bibfield  {author} {\bibinfo {author} {\bibfnamefont {Z.}~\bibnamefont
  {Aqua}}, \bibinfo {author} {\bibfnamefont {M.~S.}\ \bibnamefont {Kim}}, \
  and\ \bibinfo {author} {\bibfnamefont {B.}~\bibnamefont {Dayan}},\ }\bibfield
   {title} {\enquote {\bibinfo {title} {Generation of optical {Fock} and {W}
  states with single-atom-based bright quantum scissors},}\ }\href {\doibase
  10.1364/prj.7.000a45} {\bibfield  {journal} {\bibinfo  {journal} {Photonics
  Research}\ }\textbf {\bibinfo {volume} {7}},\ \bibinfo {pages} {A45}
  (\bibinfo {year} {2019})}\BibitemShut {NoStop}%
\bibitem [{\citenamefont {Lindner}\ and\ \citenamefont
  {Rudolph}(2009)}]{Lindner2009}%
  \BibitemOpen
  \bibfield  {author} {\bibinfo {author} {\bibfnamefont {N.~H.}\ \bibnamefont
  {Lindner}}\ and\ \bibinfo {author} {\bibfnamefont {T.}~\bibnamefont
  {Rudolph}},\ }\bibfield  {title} {\enquote {\bibinfo {title} {Proposal for
  pulsed on-demand sources of photonic cluster state strings},}\ }\href
  {\doibase 10.1103/PhysRevLett.103.113602} {\bibfield  {journal} {\bibinfo
  {journal} {Phys. Rev. Lett.}\ }\textbf {\bibinfo {volume} {103}},\ \bibinfo
  {pages} {113602} (\bibinfo {year} {2009})}\BibitemShut {NoStop}%
\bibitem [{\citenamefont {Schwartz}\ \emph {et~al.}(2016)\citenamefont
  {Schwartz}, \citenamefont {Cogan}, \citenamefont {Schmidgall}, \citenamefont
  {Don}, \citenamefont {Gantz}, \citenamefont {Kenneth}, \citenamefont
  {Lindner},\ and\ \citenamefont {Gershoni}}]{Schwartz2016}%
  \BibitemOpen
  \bibfield  {author} {\bibinfo {author} {\bibfnamefont {I.}~\bibnamefont
  {Schwartz}}, \bibinfo {author} {\bibfnamefont {D.}~\bibnamefont {Cogan}},
  \bibinfo {author} {\bibfnamefont {E.~R.}\ \bibnamefont {Schmidgall}},
  \bibinfo {author} {\bibfnamefont {Y.}~\bibnamefont {Don}}, \bibinfo {author}
  {\bibfnamefont {L.}~\bibnamefont {Gantz}}, \bibinfo {author} {\bibfnamefont
  {O.}~\bibnamefont {Kenneth}}, \bibinfo {author} {\bibfnamefont {N.~H.}\
  \bibnamefont {Lindner}}, \ and\ \bibinfo {author} {\bibfnamefont
  {D.}~\bibnamefont {Gershoni}},\ }\bibfield  {title} {\enquote {\bibinfo
  {title} {Deterministic generation of a cluster state of entangled photons},}\
  }\href {\doibase 10.1126/science.aah4758} {\bibfield  {journal} {\bibinfo
  {journal} {Science}\ }\textbf {\bibinfo {volume} {354}},\ \bibinfo {pages}
  {434} (\bibinfo {year} {2016})}\BibitemShut {NoStop}%
\bibitem [{\citenamefont {Istrati}\ \emph {et~al.}(2020)\citenamefont
  {Istrati}, \citenamefont {Pilnyak}, \citenamefont {Loredo}, \citenamefont
  {Ant{\'{o}}n}, \citenamefont {Somaschi}, \citenamefont {Hilaire},
  \citenamefont {Ollivier}, \citenamefont {Esmann}, \citenamefont {Cohen},
  \citenamefont {Vidro}, \citenamefont {Millet}, \citenamefont {Lemaitre},
  \citenamefont {Sagnes}, \citenamefont {Harouri}, \citenamefont {Lanco},
  \citenamefont {Senellart},\ and\ \citenamefont {Eisenberg}}]{Istrati2020}%
  \BibitemOpen
  \bibfield  {author} {\bibinfo {author} {\bibfnamefont {D.}~\bibnamefont
  {Istrati}}, \bibinfo {author} {\bibfnamefont {Y.}~\bibnamefont {Pilnyak}},
  \bibinfo {author} {\bibfnamefont {J.~C.}\ \bibnamefont {Loredo}}, \bibinfo
  {author} {\bibfnamefont {C.}~\bibnamefont {Ant{\'{o}}n}}, \bibinfo {author}
  {\bibfnamefont {N.}~\bibnamefont {Somaschi}}, \bibinfo {author}
  {\bibfnamefont {P.}~\bibnamefont {Hilaire}}, \bibinfo {author} {\bibfnamefont
  {H.}~\bibnamefont {Ollivier}}, \bibinfo {author} {\bibfnamefont
  {M.}~\bibnamefont {Esmann}}, \bibinfo {author} {\bibfnamefont
  {L.}~\bibnamefont {Cohen}}, \bibinfo {author} {\bibfnamefont
  {L.}~\bibnamefont {Vidro}}, \bibinfo {author} {\bibfnamefont
  {C.}~\bibnamefont {Millet}}, \bibinfo {author} {\bibfnamefont
  {A.}~\bibnamefont {Lemaitre}}, \bibinfo {author} {\bibfnamefont
  {I.}~\bibnamefont {Sagnes}}, \bibinfo {author} {\bibfnamefont
  {A.}~\bibnamefont {Harouri}}, \bibinfo {author} {\bibfnamefont
  {L.}~\bibnamefont {Lanco}}, \bibinfo {author} {\bibfnamefont
  {P.}~\bibnamefont {Senellart}}, \ and\ \bibinfo {author} {\bibfnamefont
  {H.~S.}\ \bibnamefont {Eisenberg}},\ }\bibfield  {title} {\enquote {\bibinfo
  {title} {Sequential generation of linear cluster states from a single photon
  emitter},}\ }\href {\doibase 10.1038/s41467-020-19341-4} {\bibfield
  {journal} {\bibinfo  {journal} {Nat. Commun.}\ }\textbf {\bibinfo {volume}
  {11}} (\bibinfo {year} {2020}),\ 10.1038/s41467-020-19341-4}\BibitemShut
  {NoStop}%
\bibitem [{\citenamefont {Smirnov}\ \emph {et~al.}(2017)\citenamefont
  {Smirnov}, \citenamefont {Reznychenko}, \citenamefont {Auff\`eves},\ and\
  \citenamefont {Lanco}}]{Smirnov2017}%
  \BibitemOpen
  \bibfield  {author} {\bibinfo {author} {\bibfnamefont {D.~S.}\ \bibnamefont
  {Smirnov}}, \bibinfo {author} {\bibfnamefont {B.}~\bibnamefont
  {Reznychenko}}, \bibinfo {author} {\bibfnamefont {A.}~\bibnamefont
  {Auff\`eves}}, \ and\ \bibinfo {author} {\bibfnamefont {L.}~\bibnamefont
  {Lanco}},\ }\bibfield  {title} {\enquote {\bibinfo {title} {Measurement back
  action and spin noise spectroscopy in a charged cavity qed device in the
  strong coupling regime},}\ }\href {\doibase 10.1103/PhysRevB.96.165308}
  {\bibfield  {journal} {\bibinfo  {journal} {Phys. Rev. B}\ }\textbf {\bibinfo
  {volume} {96}},\ \bibinfo {pages} {165308} (\bibinfo {year}
  {2017})}\BibitemShut {NoStop}%
\bibitem [{\citenamefont {Pichler}\ \emph {et~al.}(2017)\citenamefont
  {Pichler}, \citenamefont {Choi}, \citenamefont {Zoller},\ and\ \citenamefont
  {Lukin}}]{Pichler2017}%
  \BibitemOpen
  \bibfield  {author} {\bibinfo {author} {\bibfnamefont {H.}~\bibnamefont
  {Pichler}}, \bibinfo {author} {\bibfnamefont {S.}~\bibnamefont {Choi}},
  \bibinfo {author} {\bibfnamefont {P.}~\bibnamefont {Zoller}}, \ and\ \bibinfo
  {author} {\bibfnamefont {M.~D.}\ \bibnamefont {Lukin}},\ }\bibfield  {title}
  {\enquote {\bibinfo {title} {Universal photonic quantum computation via
  time-delayed feedback},}\ }\href {\doibase 10.1073/pnas.1711003114}
  {\bibfield  {journal} {\bibinfo  {journal} {Proceedings of the National
  Academy of Sciences}\ }\textbf {\bibinfo {volume} {114}},\ \bibinfo {pages}
  {11362--11367} (\bibinfo {year} {2017})}\BibitemShut {NoStop}%
\bibitem [{\citenamefont {Thomas}\ \emph {et~al.}(2022)\citenamefont {Thomas},
  \citenamefont {Ruscio}, \citenamefont {Morin},\ and\ \citenamefont
  {Rempe}}]{Thomas2022}%
  \BibitemOpen
  \bibfield  {author} {\bibinfo {author} {\bibfnamefont {P.}~\bibnamefont
  {Thomas}}, \bibinfo {author} {\bibfnamefont {L.}~\bibnamefont {Ruscio}},
  \bibinfo {author} {\bibfnamefont {O.}~\bibnamefont {Morin}}, \ and\ \bibinfo
  {author} {\bibfnamefont {G.}~\bibnamefont {Rempe}},\ }\bibfield  {title}
  {\enquote {\bibinfo {title} {Efficient generation of entangled multiphoton
  graph states from a single atom},}\ }\href {\doibase
  10.1038/s41586-022-04987-5} {\bibfield  {journal} {\bibinfo  {journal}
  {Nature}\ }\textbf {\bibinfo {volume} {608}},\ \bibinfo {pages} {677}
  (\bibinfo {year} {2022})}\BibitemShut {NoStop}%
\bibitem [{\citenamefont {Rupasov}\ and\ \citenamefont
  {Yudson}(1984)}]{Yudson1984}%
  \BibitemOpen
  \bibfield  {author} {\bibinfo {author} {\bibfnamefont {V.I.}\ \bibnamefont
  {Rupasov}}\ and\ \bibinfo {author} {\bibfnamefont {V.I.}\ \bibnamefont
  {Yudson}},\ }\bibfield  {title} {\enquote {\bibinfo {title} {{Exact {D}icke
  superradiance theory: {B}ethe wavefunctions in the discrete atom model}},}\
  }\href@noop {} {\bibfield  {journal} {\bibinfo  {journal} {Sov. Phys. JETP}\
  }\textbf {\bibinfo {volume} {59}},\ \bibinfo {pages} {478} (\bibinfo {year}
  {1984})}\BibitemShut {NoStop}%
\bibitem [{\citenamefont {Shen}\ and\ \citenamefont {Fan}(2007)}]{Fan2007}%
  \BibitemOpen
  \bibfield  {author} {\bibinfo {author} {\bibfnamefont {J.~T.}\ \bibnamefont
  {Shen}}\ and\ \bibinfo {author} {\bibfnamefont {S.}~\bibnamefont {Fan}},\
  }\bibfield  {title} {\enquote {\bibinfo {title} {{Strongly Correlated
  Two-Photon Transport in a One-Dimensional Waveguide Coupled to a Two-Level
  System}},}\ }\href {\doibase 10.1103/PhysRevLett.98.153003} {\bibfield
  {journal} {\bibinfo  {journal} {Phys. Rev. Lett.}\ }\textbf {\bibinfo
  {volume} {98}},\ \bibinfo {pages} {153003} (\bibinfo {year}
  {2007})}\BibitemShut {NoStop}%
\bibitem [{\citenamefont {Shi}\ and\ \citenamefont {Sun}(2009)}]{Shi2009}%
  \BibitemOpen
  \bibfield  {author} {\bibinfo {author} {\bibfnamefont {T.}~\bibnamefont
  {Shi}}\ and\ \bibinfo {author} {\bibfnamefont {C.~P.}\ \bibnamefont {Sun}},\
  }\bibfield  {title} {\enquote {\bibinfo {title}
  {{{L}ehmann-{S}ymanzik-{Z}immermann reduction approach to multiphoton
  scattering in coupled-resonator arrays}},}\ }\href {\doibase
  10.1103/PhysRevB.79.205111} {\bibfield  {journal} {\bibinfo  {journal} {Phys.
  Rev. B}\ }\textbf {\bibinfo {volume} {79}},\ \bibinfo {pages} {205111}
  (\bibinfo {year} {2009})}\BibitemShut {NoStop}%
\bibitem [{\citenamefont {Zheng}\ and\ \citenamefont
  {Baranger}(2013)}]{Baranger2013}%
  \BibitemOpen
  \bibfield  {author} {\bibinfo {author} {\bibfnamefont {H.}~\bibnamefont
  {Zheng}}\ and\ \bibinfo {author} {\bibfnamefont {H.~U.}\ \bibnamefont
  {Baranger}},\ }\bibfield  {title} {\enquote {\bibinfo {title} {{Persistent
  Quantum Beats and Long-Distance Entanglement from Waveguide-Mediated
  Interactions}},}\ }\href {\doibase 10.1103/PhysRevLett.110.113601} {\bibfield
   {journal} {\bibinfo  {journal} {Phys. Rev. Lett.}\ }\textbf {\bibinfo
  {volume} {110}},\ \bibinfo {pages} {113601} (\bibinfo {year}
  {2013})}\BibitemShut {NoStop}%
\bibitem [{\citenamefont {Xu}\ and\ \citenamefont {Fan}(2015)}]{Xu2015}%
  \BibitemOpen
  \bibfield  {author} {\bibinfo {author} {\bibfnamefont {S.}~\bibnamefont
  {Xu}}\ and\ \bibinfo {author} {\bibfnamefont {S.}~\bibnamefont {Fan}},\
  }\bibfield  {title} {\enquote {\bibinfo {title} {Input-output formalism for
  few-photon transport: A systematic treatment beyond two photons},}\ }\href
  {\doibase 10.1103/PhysRevA.91.043845} {\bibfield  {journal} {\bibinfo
  {journal} {Phys. Rev. A}\ }\textbf {\bibinfo {volume} {91}},\ \bibinfo
  {pages} {043845} (\bibinfo {year} {2015})}\BibitemShut {NoStop}%
\bibitem [{\citenamefont {Trivedi}\ \emph {et~al.}(2018)\citenamefont
  {Trivedi}, \citenamefont {Fischer}, \citenamefont {Xu}, \citenamefont {Fan},\
  and\ \citenamefont {Vuckovic}}]{Trivedi2018}%
  \BibitemOpen
  \bibfield  {author} {\bibinfo {author} {\bibfnamefont {R.}~\bibnamefont
  {Trivedi}}, \bibinfo {author} {\bibfnamefont {K.}~\bibnamefont {Fischer}},
  \bibinfo {author} {\bibfnamefont {S.}~\bibnamefont {Xu}}, \bibinfo {author}
  {\bibfnamefont {S.}~\bibnamefont {Fan}}, \ and\ \bibinfo {author}
  {\bibfnamefont {J.}~\bibnamefont {Vuckovic}},\ }\bibfield  {title} {\enquote
  {\bibinfo {title} {Few-photon scattering and emission from low-dimensional
  quantum systems},}\ }\href {\doibase 10.1103/PhysRevB.98.144112} {\bibfield
  {journal} {\bibinfo  {journal} {Phys. Rev. B}\ }\textbf {\bibinfo {volume}
  {98}},\ \bibinfo {pages} {144112} (\bibinfo {year} {2018})}\BibitemShut
  {NoStop}%
\bibitem [{\citenamefont {Shi}\ \emph {et~al.}(2015)\citenamefont {Shi},
  \citenamefont {Chang},\ and\ \citenamefont {Cirac}}]{Shi2015}%
  \BibitemOpen
  \bibfield  {author} {\bibinfo {author} {\bibfnamefont {T.}~\bibnamefont
  {Shi}}, \bibinfo {author} {\bibfnamefont {D.~E.}\ \bibnamefont {Chang}}, \
  and\ \bibinfo {author} {\bibfnamefont {J.~I.}\ \bibnamefont {Cirac}},\
  }\bibfield  {title} {\enquote {\bibinfo {title} {Multiphoton-scattering
  theory and generalized master equations},}\ }\href {\doibase
  10.1103/PhysRevA.92.053834} {\bibfield  {journal} {\bibinfo  {journal} {Phys.
  Rev. A}\ }\textbf {\bibinfo {volume} {92}},\ \bibinfo {pages} {053834}
  (\bibinfo {year} {2015})}\BibitemShut {NoStop}%
\bibitem [{\citenamefont {Gonz\'alez-Tudela}\ \emph {et~al.}(2015)\citenamefont
  {Gonz\'alez-Tudela}, \citenamefont {Paulisch}, \citenamefont {Chang},
  \citenamefont {Kimble},\ and\ \citenamefont {Cirac}}]{Tudela2015}%
  \BibitemOpen
  \bibfield  {author} {\bibinfo {author} {\bibfnamefont {A.}~\bibnamefont
  {Gonz\'alez-Tudela}}, \bibinfo {author} {\bibfnamefont {V.}~\bibnamefont
  {Paulisch}}, \bibinfo {author} {\bibfnamefont {D.~E.}\ \bibnamefont {Chang}},
  \bibinfo {author} {\bibfnamefont {H.~J.}\ \bibnamefont {Kimble}}, \ and\
  \bibinfo {author} {\bibfnamefont {J.~I.}\ \bibnamefont {Cirac}},\ }\bibfield
  {title} {\enquote {\bibinfo {title} {Deterministic generation of arbitrary
  photonic states assisted by dissipation},}\ }\href {\doibase
  10.1103/PhysRevLett.115.163603} {\bibfield  {journal} {\bibinfo  {journal}
  {Phys. Rev. Lett.}\ }\textbf {\bibinfo {volume} {115}},\ \bibinfo {pages}
  {163603} (\bibinfo {year} {2015})}\BibitemShut {NoStop}%
\bibitem [{\citenamefont {Caneva}\ \emph {et~al.}(2015)\citenamefont {Caneva},
  \citenamefont {Manzoni}, \citenamefont {Shi}, \citenamefont {Douglas},
  \citenamefont {Cirac},\ and\ \citenamefont {Chang}}]{Caneva2015}%
  \BibitemOpen
  \bibfield  {author} {\bibinfo {author} {\bibfnamefont {T.}~\bibnamefont
  {Caneva}}, \bibinfo {author} {\bibfnamefont {M.~T.}\ \bibnamefont {Manzoni}},
  \bibinfo {author} {\bibfnamefont {T.}~\bibnamefont {Shi}}, \bibinfo {author}
  {\bibfnamefont {J.~S.}\ \bibnamefont {Douglas}}, \bibinfo {author}
  {\bibfnamefont {J.~I.}\ \bibnamefont {Cirac}}, \ and\ \bibinfo {author}
  {\bibfnamefont {D.~E.}\ \bibnamefont {Chang}},\ }\bibfield  {title} {\enquote
  {\bibinfo {title} {Quantum dynamics of propagating photons with strong
  interactions: a generalized input{\textendash}output formalism},}\ }\href
  {\doibase 10.1088/1367-2630/17/11/113001} {\bibfield  {journal} {\bibinfo
  {journal} {New J. Phys.}\ }\textbf {\bibinfo {volume} {17}},\ \bibinfo
  {pages} {113001} (\bibinfo {year} {2015})}\BibitemShut {NoStop}%
\bibitem [{\citenamefont {Fang}\ and\ \citenamefont
  {Baranger}(2016)}]{Fang2016}%
  \BibitemOpen
  \bibfield  {author} {\bibinfo {author} {\bibfnamefont {Y.~L.}\ \bibnamefont
  {Fang}}\ and\ \bibinfo {author} {\bibfnamefont {H.~U.}\ \bibnamefont
  {Baranger}},\ }\bibfield  {title} {\enquote {\bibinfo {title} {Reprint of :
  Photon correlations generated by inelastic scattering in a one-dimensional
  waveguide coupled to three-level systems},}\ }\href {\doibase
  10.1016/j.physe.2016.02.015} {\bibfield  {journal} {\bibinfo  {journal}
  {Physica E}\ }\textbf {\bibinfo {volume} {82}},\ \bibinfo {pages} {71}
  (\bibinfo {year} {2016})}\BibitemShut {NoStop}%
\bibitem [{\citenamefont {Iversen}\ and\ \citenamefont
  {Pohl}(2021)}]{Iversen2021}%
  \BibitemOpen
  \bibfield  {author} {\bibinfo {author} {\bibfnamefont {O.~A.}\ \bibnamefont
  {Iversen}}\ and\ \bibinfo {author} {\bibfnamefont {T.}~\bibnamefont {Pohl}},\
  }\bibfield  {title} {\enquote {\bibinfo {title} {Strongly correlated states
  of light and repulsive photons in chiral chains of three-level quantum
  emitters},}\ }\href {\doibase 10.1103/PhysRevLett.126.083605} {\bibfield
  {journal} {\bibinfo  {journal} {Phys. Rev. Lett.}\ }\textbf {\bibinfo
  {volume} {126}},\ \bibinfo {pages} {083605} (\bibinfo {year}
  {2021})}\BibitemShut {NoStop}%
\bibitem [{\citenamefont {Witthaut}\ and\ \citenamefont
  {S{\o}rensen}(2010)}]{Witthaut2010}%
  \BibitemOpen
  \bibfield  {author} {\bibinfo {author} {\bibfnamefont {D.}~\bibnamefont
  {Witthaut}}\ and\ \bibinfo {author} {\bibfnamefont {A.~S.}\ \bibnamefont
  {S{\o}rensen}},\ }\bibfield  {title} {\enquote {\bibinfo {title} {Photon
  scattering by a three-level emitter in a one-dimensional waveguide},}\ }\href
  {\doibase 10.1088/1367-2630/12/4/043052} {\bibfield  {journal} {\bibinfo
  {journal} {New J. Phys.}\ }\textbf {\bibinfo {volume} {12}},\ \bibinfo
  {pages} {043052} (\bibinfo {year} {2010})}\BibitemShut {NoStop}%
\bibitem [{\citenamefont {Li}\ \emph {et~al.}(2012)\citenamefont {Li},
  \citenamefont {Aolita}, \citenamefont {Chang},\ and\ \citenamefont
  {Kwek}}]{Li2012}%
  \BibitemOpen
  \bibfield  {author} {\bibinfo {author} {\bibfnamefont {Y.}~\bibnamefont
  {Li}}, \bibinfo {author} {\bibfnamefont {L.}~\bibnamefont {Aolita}}, \bibinfo
  {author} {\bibfnamefont {Darrick~E.}\ \bibnamefont {Chang}}, \ and\ \bibinfo
  {author} {\bibfnamefont {L.~C.}\ \bibnamefont {Kwek}},\ }\bibfield  {title}
  {\enquote {\bibinfo {title} {Robust-fidelity atom-photon entangling gates in
  the weak-coupling regime},}\ }\href {\doibase 10.1103/PhysRevLett.109.160504}
  {\bibfield  {journal} {\bibinfo  {journal} {Phys. Rev. Lett.}\ }\textbf
  {\bibinfo {volume} {109}},\ \bibinfo {pages} {160504} (\bibinfo {year}
  {2012})}\BibitemShut {NoStop}%
\bibitem [{\citenamefont {Bradford}\ and\ \citenamefont
  {Shen}(2012)}]{Bradford2012}%
  \BibitemOpen
  \bibfield  {author} {\bibinfo {author} {\bibfnamefont {M.}~\bibnamefont
  {Bradford}}\ and\ \bibinfo {author} {\bibfnamefont {J.-T.}\ \bibnamefont
  {Shen}},\ }\bibfield  {title} {\enquote {\bibinfo {title} {Single-photon
  frequency conversion by exploiting quantum interference},}\ }\href {\doibase
  10.1103/PhysRevA.85.043814} {\bibfield  {journal} {\bibinfo  {journal} {Phys.
  Rev. A}\ }\textbf {\bibinfo {volume} {85}},\ \bibinfo {pages} {043814}
  (\bibinfo {year} {2012})}\BibitemShut {NoStop}%
\bibitem [{\citenamefont {Martens}\ \emph {et~al.}(2013)\citenamefont
  {Martens}, \citenamefont {Longo},\ and\ \citenamefont {Busch}}]{Martens2013}%
  \BibitemOpen
  \bibfield  {author} {\bibinfo {author} {\bibfnamefont {C.}~\bibnamefont
  {Martens}}, \bibinfo {author} {\bibfnamefont {P.}~\bibnamefont {Longo}}, \
  and\ \bibinfo {author} {\bibfnamefont {K.}~\bibnamefont {Busch}},\ }\bibfield
   {title} {\enquote {\bibinfo {title} {Photon transport in one-dimensional
  systems coupled to three-level quantum impurities},}\ }\href {\doibase
  10.1088/1367-2630/15/8/083019} {\bibfield  {journal} {\bibinfo  {journal}
  {New J. Phys.}\ }\textbf {\bibinfo {volume} {15}},\ \bibinfo {pages} {083019}
  (\bibinfo {year} {2013})}\BibitemShut {NoStop}%
\bibitem [{\citenamefont {Das}\ \emph {et~al.}(2018)\citenamefont {Das},
  \citenamefont {Elfving}, \citenamefont {Reiter},\ and\ \citenamefont
  {S\o{}rensen}}]{Das2018}%
  \BibitemOpen
  \bibfield  {author} {\bibinfo {author} {\bibfnamefont {S.}~\bibnamefont
  {Das}}, \bibinfo {author} {\bibfnamefont {V.~E.}\ \bibnamefont {Elfving}},
  \bibinfo {author} {\bibfnamefont {F.n}\ \bibnamefont {Reiter}}, \ and\
  \bibinfo {author} {\bibfnamefont {A.~S.}\ \bibnamefont {S\o{}rensen}},\
  }\bibfield  {title} {\enquote {\bibinfo {title} {Photon scattering from a
  system of multilevel quantum emitters. ii. application to emitters coupled to
  a one-dimensional waveguide},}\ }\href {\doibase 10.1103/PhysRevA.97.043838}
  {\bibfield  {journal} {\bibinfo  {journal} {Phys. Rev. A}\ }\textbf {\bibinfo
  {volume} {97}},\ \bibinfo {pages} {043838} (\bibinfo {year}
  {2018})}\BibitemShut {NoStop}%
\bibitem [{\citenamefont {Zhong}\ \emph {et~al.}(2023)\citenamefont {Zhong},
  \citenamefont {Rituraj}, \citenamefont {Dinc},\ and\ \citenamefont
  {Fan}}]{Zhong2023}%
  \BibitemOpen
  \bibfield  {author} {\bibinfo {author} {\bibfnamefont {J.}~\bibnamefont
  {Zhong}}, \bibinfo {author} {\bibnamefont {Rituraj}}, \bibinfo {author}
  {\bibfnamefont {F.}~\bibnamefont {Dinc}}, \ and\ \bibinfo {author}
  {\bibfnamefont {S.}~\bibnamefont {Fan}},\ }\bibfield  {title} {\enquote
  {\bibinfo {title} {Detecting the relative phase between different frequency
  components of a photon using a three-level $\mathrm{\ensuremath{\Lambda}}$
  atom coupled to a waveguide},}\ }\href {\doibase
  10.1103/PhysRevA.107.L051702} {\bibfield  {journal} {\bibinfo  {journal}
  {Phys. Rev. A}\ }\textbf {\bibinfo {volume} {107}},\ \bibinfo {pages}
  {L051702} (\bibinfo {year} {2023})}\BibitemShut {NoStop}%
\bibitem [{\citenamefont {Zhang}\ and\ \citenamefont {Yang}(2023)}]{Zhang2023}%
  \BibitemOpen
  \bibfield  {author} {\bibinfo {author} {\bibfnamefont {Z.~L.}\ \bibnamefont
  {Zhang}}\ and\ \bibinfo {author} {\bibfnamefont {L.~P.}\ \bibnamefont
  {Yang}},\ }\bibfield  {title} {\enquote {\bibinfo {title} {Limits of
  single-photon storage in a single $\mathrm{\ensuremath{\Lambda}}$-type
  atom},}\ }\href {\doibase 10.1103/PhysRevA.107.063704} {\bibfield  {journal}
  {\bibinfo  {journal} {Phys. Rev. A}\ }\textbf {\bibinfo {volume} {107}},\
  \bibinfo {pages} {063704} (\bibinfo {year} {2023})}\BibitemShut {NoStop}%
\bibitem [{\citenamefont {Chan}\ \emph {et~al.}(2023)\citenamefont {Chan},
  \citenamefont {Tiranov}, \citenamefont {Appel}, \citenamefont {Wang},
  \citenamefont {Midolo}, \citenamefont {Scholz}, \citenamefont {Wieck},
  \citenamefont {Ludwig}, \citenamefont {S{\o}rensen},\ and\ \citenamefont
  {Lodahl}}]{Chan2023}%
  \BibitemOpen
  \bibfield  {author} {\bibinfo {author} {\bibfnamefont {M.~L.}\ \bibnamefont
  {Chan}}, \bibinfo {author} {\bibfnamefont {A.}~\bibnamefont {Tiranov}},
  \bibinfo {author} {\bibfnamefont {M.~H.}\ \bibnamefont {Appel}}, \bibinfo
  {author} {\bibfnamefont {Y.}~\bibnamefont {Wang}}, \bibinfo {author}
  {\bibfnamefont {L.}~\bibnamefont {Midolo}}, \bibinfo {author} {\bibfnamefont
  {S.}~\bibnamefont {Scholz}}, \bibinfo {author} {\bibfnamefont {A.~D.}\
  \bibnamefont {Wieck}}, \bibinfo {author} {\bibfnamefont {A.}~\bibnamefont
  {Ludwig}}, \bibinfo {author} {\bibfnamefont {A.~S.}\ \bibnamefont
  {S{\o}rensen}}, \ and\ \bibinfo {author} {\bibfnamefont {P.}~\bibnamefont
  {Lodahl}},\ }\bibfield  {title} {\enquote {\bibinfo {title} {On-chip
  spin-photon entanglement based on photon-scattering of a quantum dot},}\
  }\href {\doibase 10.1038/s41534-023-00717-5} {\bibfield  {journal} {\bibinfo
  {journal} {npj Quantum Information}\ }\textbf {\bibinfo {volume} {9}}
  (\bibinfo {year} {2023}),\ 10.1038/s41534-023-00717-5}\BibitemShut {NoStop}%
\bibitem [{Note1()}]{Note1}%
  \BibitemOpen
  \bibinfo {note} {See Supplementary for for the simplified expression for
  $|S^{XXx\leftarrow XXx}|^2$ that matches the well-known result for a
  two-level atom~\cite {Fan2007}}\BibitemShut {NoStop}%
\bibitem [{\citenamefont {Yang}\ and\ \citenamefont {Cao}(2004)}]{Yang2004}%
  \BibitemOpen
  \bibfield  {author} {\bibinfo {author} {\bibfnamefont {M.}~\bibnamefont
  {Yang}}\ and\ \bibinfo {author} {\bibfnamefont {Z.L.}\ \bibnamefont {Cao}},\
  }\bibfield  {title} {\enquote {\bibinfo {title} {Entanglement distillation
  for w class states},}\ }\href {\doibase 10.1016/j.physa.2004.02.016}
  {\bibfield  {journal} {\bibinfo  {journal} {Physica A}\ }\textbf {\bibinfo
  {volume} {337}},\ \bibinfo {pages} {141} (\bibinfo {year}
  {2004})}\BibitemShut {NoStop}%
\bibitem [{Note2()}]{Note2}%
  \BibitemOpen
  \bibinfo {note} {See Supplementary for the explicit expressions for the
  scattering matrix elements}\BibitemShut {NoStop}%
\bibitem [{Note3()}]{Note3}%
  \BibitemOpen
  \bibinfo {note} {For monochromatic three-photon incident pulse calculation
  yields $\langle P_{W_4}\rangle _\omega = 4 |t(\omega )|^4\protect \,{\protect
  \rm min}\{|t(\omega )|^2, |s(\omega )|^2\}$}\BibitemShut {NoStop}%
\end{thebibliography}%

\newpage
\onecolumngrid
\section{Supplementary information}

\subsection{Two-photon scattering by $\Lambda$-atom vs. two-level atom}

The scattering matrix element Eq.~\eqref{Eq:Sx} after collecting all the terms is simplified to 
\begin{align}
S^{XXx\leftarrow XXx}_{\omega_1',\omega_2'\leftarrow\omega_1,\omega_2} &=   t(\omega_1)t(\omega_2)  (2\pi)^2 [\delta(\omega_1-\omega_1')\delta(\omega_2-\omega_2')+ \delta(\omega_1-\omega_2')\delta(\omega_2-\omega_1')]  \\ \nonumber
&+
 \frac{2\rmi\Gamma_0^2(\omega_1+\omega_2-2\omega_0+2\rmi\Gamma_0)}{(\omega_1-\omega_0+\rmi\Gamma_0)(\omega_2-\omega_0+\rmi\Gamma_0)(\omega_1'-\omega_0+\rmi\Gamma_0)(\omega_2'-\omega_0+\rmi\Gamma_0)}  2\pi \delta(\omega_1+\omega_2-\omega_1'-\omega_2')
\,.
\end{align}
The singularities at $\omega_{1',2'} \to \omega_{1',2'}$ that seem to be present in Eq.~\eqref{Eq:Sx} are in fact cancelled. Note that it is not the case for $S^{XYy\leftarrow XXx}$, Eq.~\eqref{Eq:Sy},  that is not symmetrized with respect to $\omega_1',\omega_2'$ and retains the singularities. 

The first line describes the coherent transmission of independent photons and the second line is the elastic  two-photon scattering. The latter term appears to be 4 times smaller than the corresponding scattering amplitude in a chiral waveguide with a two-level atom, and is exactly the same as for the two-level atom in a bi-directional waveguide~\cite{Yudson1984,Fan2007,Sheremet2023}.

\subsection{Two-photon scattering matrix in frequency-time domain}

We take the scattering matrix in the frequency domain Eqs.~\eqref{Eq:Sy}--\eqref{Eq:Sx} and perform Fourier transform over the final frequencies $\omega'$ and $\omega_2'$. The result is 
\begin{align}
S^{XXx\leftarrow XXx}_{t_1,t_2 \leftarrow \omega_1,\omega_2} 
&= \e^{-\rmi\omega_1t_1-\rmi\omega_2t_2}  + (\omega_1\leftrightarrow \omega_2)\\\nonumber
&+s(\omega_2)\e^{-\rmi\omega_1t_1-\rmi\omega_2t_2}  + (\omega_1\leftrightarrow \omega_2)+ (t_1\leftrightarrow t_2)+ (\omega_1\leftrightarrow \omega_2, t_1\leftrightarrow t_2)\nonumber\\
&+s(\omega_1)s(\omega_2) \theta(t_2-t_1) \left[ \e^{-\rmi\omega_1t_1-\rmi\omega_2t_2}-\e^{-\Gamma_0(t_2-t_1)-\rmi (\omega_1+\omega_2) t_1} \right] 
+  (\omega_1\leftrightarrow \omega_2)+ (t_1\leftrightarrow t_2)+ (\omega_1\leftrightarrow \omega_2, t_1\leftrightarrow t_2) \nonumber\\ 
=& 2\left[ t(\omega_1)t(\omega_2)\cos \frac{(\omega_2-\omega_1)(t_2-t_1)}{2} - s(\omega_1)s(\omega_2)\e^{-(\Gamma_0-\rmi\frac{\omega_1+\omega_2}{2})|t_2-t_1|}
\right] \e^{-\rmi (\omega_1+\omega_2) (t_1+t_2)/2}
\end{align}
\begin{align}
S^{XYy\leftarrow XXx}_{t_1,t_2 \leftarrow\omega_1',\omega_2'} 
&=s(\omega_2)\e^{-\rmi\omega_1t_1-\rmi\omega_2t_2}  + (\omega_1\leftrightarrow \omega_2)\\\nonumber
&+s(\omega_1)s(\omega_2) \theta(t_2-t_1) \left[ \e^{-\rmi\omega_1t_1-\rmi\omega_2t_2}-\e^{-\Gamma_0(t_2-t_1)-\rmi (\omega_1+\omega_2) t_1} \right] 
+  (\omega_1\leftrightarrow \omega_2) \\
=& \Big\{\theta(t_2-t_1) \left[ t(\omega_1)s(\omega_2)\e^{\frac{-\rmi (\omega_2-\omega_1)(t_2-t_1)}{2}} +
t(\omega_2)s(\omega_1)\e^{\frac{\rmi (\omega_2-\omega_1)(t_2-t_1)}{2}}
-2 s(\omega_1)s(\omega_2)\e^{-(\Gamma_0-\rmi\frac{\omega_1+\omega_2}{2})|t_2-t_1|}
\right] \nonumber\\
&+
\theta(t_1-t_2) \left[ s(\omega_2')\e^{\frac{-\rmi (\omega_2-\omega_1)(t_2-t_1)}{2}} +
s(\omega_1)\e^{\frac{\rmi (\omega_2-\omega_1)(t_2-t_1)}{2}}
\right] \Big\}
\e^{-\rmi (\omega_1+\omega_2) (t_1+t_2)/2}
\end{align}

\subsection{There-photon scattering matrix}

Evaluation of diagrams in Fig.~\ref{fig:dia}(d) yields the three-photon scattering matrix in the frequency domain:
\begin{align}
    S^{XXXx\leftarrow XXXx}_{\omega_1',\omega_2',\omega_3' \leftarrow \omega_1,\omega_2,\omega_3} 
    &=
\frac{-\rmi\Gamma_0^2 \,\, 2\pi\delta(\omega_1'+\omega_2'+\omega_3'-\omega_1-\omega_2-\omega_3)}{(\omega_1-\omega_0+\rmi\Gamma_0)(\omega_1-\omega_1'+\rmi 0)(\omega_1+\omega_2-\omega_1'-\omega_0+\rmi\Gamma_0)(\omega_3'-\omega_3+\rmi 0)(\omega_3'-\omega_0+\rmi \Gamma_0)} \\ \nonumber
&\qquad+ \text{[permutations of $(\omega_1',\omega_2',\omega_3')$ and $ (\omega_1,\omega_2,\omega_3)$]} \\
&+
\frac{-\rmi\Gamma_0^2 \,\, (2\pi)^2 \delta(\omega_1'+\omega_2'-\omega_1-\omega_2)\delta(\omega_3'-\omega_3)}{(\omega_1-\omega_0+\rmi\Gamma_0)(\omega_1-\omega_1'+\rmi 0)(\omega_2'-\omega_0+\rmi\Gamma_0)} 
\\ \nonumber
&\qquad+ \text{[permutations of $(\omega_1',\omega_2',\omega_3')$ and $ (\omega_1,\omega_2,\omega_3)$]} \\
&+ [1+s(\omega_1')+ s(\omega_2')+ s(\omega_3')] (2\pi)^3 \delta(\omega_1'-\omega_1)\delta(\omega_2'-\omega_2)\delta(\omega_3'-\omega_3) \\\nonumber
&\qquad+ \text{[permutations of $ (\omega_1,\omega_2,\omega_3)$]} 
\end{align}
\begin{align}
    S^{XXYy \leftarrow  XXXx}_{\omega_1',\omega_2',\omega_3' \leftarrow \omega_1,\omega_2,\omega_3}
    &=
\frac{-\rmi\Gamma_0^2 \,\, 2\pi\delta(\omega_1'+\omega_2'+\omega_3'-\omega_1-\omega_2-\omega_3)}{(\omega_1-\omega_0+\rmi\Gamma_0)(\omega_1-\omega_1'+\rmi 0)(\omega_1+\omega_2-\omega_1'-\omega_0+\rmi\Gamma_0)(\omega_3'-\omega_3+\rmi 0)(\omega_3'-\omega_0+\rmi \Gamma_0)} \\ \nonumber
&\qquad+ \text{[permutations of $(\omega_1',\omega_2')$ and $ (\omega_1,\omega_2,\omega_3)$]} \\
&+
\frac{-\rmi\Gamma_0^2 \,\, (2\pi)^2 \delta(\omega_2'+\omega_3'-\omega_2-\omega_3)\delta(\omega_1'-\omega_1)}{(\omega_2-\omega_0+\rmi\Gamma_0)(\omega_2-\omega_2'+\rmi 0)(\omega_3'-\omega_0+\rmi\Gamma_0)} 
\\ \nonumber
&\qquad+ \text{[permutations of $(\omega_1',\omega_2')$ and $ (\omega_1,\omega_2,\omega_3)$]} \\
&+  s(\omega_3') (2\pi)^3 \delta(\omega_1'-\omega_1)\delta(\omega_2'-\omega_2)\delta(\omega_3'-\omega_3) \\\nonumber
&\qquad+ \text{[permutations of $ (\omega_1,\omega_2,\omega_3)$]} 
\end{align}
Here, the first contributions stand for the irreducible part of the scattering amplitude, the second contributions reduce to the two-photon scattering and the other photon passing the  system without interaction, and the third  contributions correspond to single-photon scattering and the other two photons passing without interaction.

Performing Fourier transform over $\omega_1',\omega_2',\omega_3'$ we get
\begin{align}
    S^{XXXx \leftarrow XXXx}_{t_1,t_2,t_3 \leftarrow  \omega_1',\omega_2',\omega_3'}
    &=
s(\omega_1)s(\omega_2)s(\omega_3)(1-\e^{(\rmi\omega_2-\Gamma_0)(t_{(2)}-t_{(1)})})(1-\e^{(\rmi\omega_3-\Gamma_0)(t_{(3)}-t_{(2)})})
\e^{-\rmi(\omega_1 t_{(1)}+\omega_2 t_{(2)}+\omega_3't_{(3)})}
\nonumber \\
&+
s(\omega_1)s(\omega_2) (\e^{-\rmi\omega_1t_{(1)}-\rmi\omega_2t_{(2)}}-\e^{-\rmi\omega_1t_{(1)}-\rmi\omega_2t_{(1)}-\Gamma_0(t_{(2)}-t_{(1)})})
\e^{-\rmi\omega_3 t_{(3)}}
\\ \nonumber
&+
s(\omega_2)s(\omega_3) (\e^{-\rmi\omega_2t_{(2)}-\rmi\omega_3t_{(3)}}-\e^{-\rmi\omega_2t_{(2)}-\rmi\omega_3t_{(2)}-\Gamma_0(t_{(3)}-t_{(2)})})
\e^{-\rmi\omega_1 t_{(1)}}
\\ 
&+
s(\omega_1)s(\omega_3) (\e^{-\rmi\omega_1t_{(1)}-\rmi\omega_3t_{(3)}}-\e^{-\rmi\omega_1t_{(1)}-\rmi\omega_3t_{(1)}-\Gamma_0(t_{(3)}-t_{(1)})})
\e^{-\rmi\omega_2 t_{(2)}}
\\ 
&+ [1+s(\omega_1)+ s(\omega_2)+ s(\omega_3)] \e^{-\rmi(\omega_1 t_{(1)}+\omega_2 t_{(2)}+\omega_3 t_{(3)})}
\\\nonumber
&\qquad+ \text{[permutations of $ (\omega_1,\omega_2,\omega_3)$]} 
\end{align}
\begin{align}
    S^{XXYy\leftarrow XXXx}_{t_1,t_2,t_3 \leftarrow  \omega_1',\omega_2',\omega_3'} 
    &=
\theta(t_3-t_>) s(\omega_1)s(\omega_2)s(\omega_3)(1-\e^{(\rmi\omega_2-\Gamma_0)(t_{(2)}-t_{(1)})})(1-\e^{(\rmi\omega_3-\Gamma_0)(t_{(3)}-t_{(2)})})
\e^{-\rmi(\omega_1 t_{(1)}+\omega_2 t_{(2)}+\omega_3 t_{(3)})}
\nonumber \\
&+
\theta(t_3-t_>) s(\omega_2)s(\omega_3) (\e^{-\rmi\omega_2t_>-\rmi\omega_3t_3}-\e^{-\rmi\omega_2t_>-\rmi\omega_3t_>-\Gamma_0(t_3-t_>)})
\e^{-\rmi\omega_1' t_<}
\\ 
&+
\theta(t_3-t_<) s(\omega_1)s(\omega_3) (\e^{-\rmi\omega_1t_<-\rmi\omega_3t_3}-\e^{-\rmi\omega_1t_<-\rmi\omega_3t_<-\Gamma_0(t_3-t_<)})
\e^{-\rmi\omega_2 t_>}
\\ 
&+ s(\omega_3) \e^{-\rmi(\omega_1 t_<+\omega_2 t_>+\omega_3 t_3}
\\\nonumber
&\qquad+ \text{[permutations of $ (\omega_1,\omega_2,\omega_3)$]} 
\end{align}
Here, $t_{(1)} \leq t_{(2)}\leq t_{(3)}$ are the sorted values of $t_1,t_2,t_3$, $t_< = {\rm min}(t_1,t_2)$, $t_> = {\rm max}(t_1,t_2)$.

\subsection{There-photon pulse transmission}

We consider excitation of the atom in the state $x$ by a three identical $X$-polarized photons which is described by the initial state wavefunction
\begin{align}
\psi^{\rm (in)}_{t_1,t_2,t_3} = \phi^{(0)}_{t_1}\phi^{(0)}_{t_2}\phi^{(0)}_{t_3} |XXx\rangle
\end{align}
If the three transmitted photons were detected at the times $t_1$, $t_2$, and $t_3$, the polarization state of the system reads
\begin{align}
\psi^{\rm (out)}_{t_1,t_2,t_3} = 
\psi^{XXXx}_{t_1,t_2,t_3} |XXXx\rangle +
\psi^{XXYy}_{t_1,t_2,t_3} |XXYy\rangle +
\psi^{XXYy}_{t_1,t_3,t_2} |XYXy\rangle +
\psi^{XXYy}_{t_3,t_1,t_2} |YXXy\rangle
\end{align}
where $\psi^{XXXx}_{t_1,t_2,t_3} $ and $\psi^{XXYy}_{t_1,t_2,t_3}$ are given by Eqs.~\eqref{Eq:XXXx}--\eqref{Eq:XXYy} of the main text.

To quantify the entanglement, we compute
\begin{align}
    &P_{W_4}(t_1,t_2,t_3) = \frac{4\, {\rm Min}\,[|\psi_{XXXx}(t_1,t_2,t_3)|^2, |\psi_{XXYy}(t_1,t_2,t_3)|^2, |\psi_{XXYy}(t_3,t_2,t_1)|^2, |\psi_{XXYy}(t_1,t_3,t_3)|^2]}{|\psi_{XXXx}(t_1,t_2,t_3)|^2+|\psi_{XXYy}(t_1,t_2,t_3)|^2+ |\psi_{XXYy}(t_3,t_2,t_1)|^2+ |\psi_{XXYy}(t_1,t_3,t_2)|^2}\\
    &\langle P_{W_4} \rangle = \frac{4 \iiint {\rm Min}\,[|\psi_{XXXx}(t_1,t_2,t_3)|^2, |\psi_{XXYy}(t_1,t_2,t_3)|^2, |\psi_{XXYy}(t_3,t_2,t_1)|^2, |\psi_{XXYy}(t_1,t_3,t_2)|^2] \,dt_1dt_2 }{\iiint [|\psi_{XXXx}(t_1,t_2,t_3)|^2+3|\psi_{XXYy}(t_1,t_2,t_3)|^2] \,dt_1dt_2 dt_3}
\end{align}

Note that from the unitarity of the scattering matrix, it follows that
\begin{align}
\iiint \left(|\psi^{XXX}_{t_1,t_2,t_3}|^2 + 3 |\psi^{XXY}_{t_1,t_2,t_3}|^2 \right) \,dt_1 dt_2 dt_3 = \left( \int |\phi^{(0)}_t|^2 dt \right)^2
\end{align}
which we have also checked numerically.

\subsection{Pulse shape optimization}

To maximize the entanglement of the scattered photons, we perform an optimization of the incident pulse shape. Namely, we take 
\begin{align}\label{Eq:Sf0}
    \phi^{(0)}_t = \left[ 1+ \sum_{n=2}^\infty (a_n + \rmi b_n)H_n(\gamma t)\right]\e^{-\rmi\omega t - \gamma^2 t^2/2}
\end{align}
where $a_n$ (for $n=3,4,...$) and $b_n$ (for $n=2,3,4,...$) are the new free parameters. We fix $a_1=b_1=a_2=0$. They would correspond to variation of pulse arrival time (that does not affect transmission), central frequency and width (that we vary using the parameters $\omega$ and $\gamma$).  
We substitute the pulse Eq.~\eqref{Eq:Sf0} into the Eqs.~\eqref{Eq:Sy}--\eqref{Eq:Sx},\eqref{Eq:XXXx}--\eqref{Eq:XXYy} to calculate the transmitted pulse. 

Finally, the values of $\langle P_{W_{3,4}}\rangle$ are calculated numerically and maximized using the conjugate gradient method. 
The results are:
\begin{center}
\begin{tabular}{c|c|c}
 & $\langle P_{W_{3}}\rangle$ & $\langle P_{W_{4}}\rangle$ \\
  & 0.8018 & 0.6237\\
   \hline
 $\omega$ & 0.8984 & 0.6747\\
 $\gamma$ & 1.0143 & 1.2721\\ 
 $b_2$ & 0.0294 & 0.0373\\
 $a_3$ & 0.0062 & 0.0021\\
 $b_3$ & 0.0147 & 0.0130\\
 $a_4$ & 0.0024 & 0.0050%\\
 %b_4 & & 
\end{tabular}
\end{center}

\end{document}